# Ultra-high-speed line-scan Raman imaging

**Qingyi Wu[1,2,3,#], Xusheng Tang[1,2,3,#], Francesco Masia[5], Peng Liang[1,2,3,4], Hao Peng[1,2,3], Lindong Shang[1,2,3],Yuntong Wang[1,2,3], Yue Qu[4], Wolfgang Langbein[6,*] and Bei Li[1,2,3,*]**

[1]Changchun Institute of Optics, Fine Mechanics and Physics, Chinese Academy of Sciences, Changchun 130033, P. R. China

[2]University of Chinese Academy of Sciences, Beijing 100049, P. R. China

[3]State Key Laboratory of Advanced Manufacturing for Optical Systems, Changchun 130033, P. R. China

[4]Haining High-tech Research Institute, Jiaxing, Zhejiang, 314408, PR China.

[5]School of Biosciences, Cardiff University, Cardiff, CF10 3AX, UK

[6]School of Physics and Astronomy, Cardiff University, Cardiff, CF24 3AA, UK

*** Correspondence:**
Wolfgang Langbein and Bei Li
langbeinww@cardiff.ac.uk and beili@ciomp.ac.cn

**[#] These authors contributed equally: Qingyi Wu, Xusheng Tang. Qingyi Wu and Xusheng Tang are the co-first authors.**

**Abstract:** Raman spectroscopic imaging has emerged as a potent tool due to its non-invasive nature and capability for chemical composition analysis. Line-scan Raman spectroscopy accelerates imaging speed by two orders of magnitude compared to point detection Raman methods. However, further enhancements in imaging speed were constrained by the readout speed of typically used charge-coupled device (CCD) spectroscopic detectors. We developed an ultra-fast line-scan Raman imaging technique based on recently available complementary metal-oxide-semiconductor (CMOS) detectors with low cost and read noise, and fast readout during exposure combined with a global shutter. Employing a high-efficiency transmission-grating imaging spectrometer, we demonstrate imaging speeds up to two orders of magnitude faster than traditional line scan Raman imaging techniques and up to four orders of magnitude faster than point scan Raman methods, achieving Raman imaging up to 80 kHz spectral rate. We demonstrate that this technology is applicable to a variety of samples, including microplastics, biological cells, and tablets, creating images in an extremely short time frame, showcasing exceptional detection capabilities and the ability to reveal detailed information.

**Key words:** Line scan Raman spectroscopy, fast imaging, microplastics, cells, tablets

## Introduction

Raman spectroscopy, a non-invasive analytical method, provides chemical composition and structural information by measuring the frequency shift of scattered light induced by molecular vibrations. Raman imaging is a widely utilized technique in fields such as material science[1-4], environmental monitoring[5-9], biology[10-14], and pharmaceutical research[15-17]. It provides spatial distribution maps of chemical components within a given sample. While spontaneous Raman spectroscopy is a widely used method, it has some inherent drawbacks, such as low signal and correspondingly slow imaging speed. Advancements in techniques such as stimulated Raman scattering (SRS) [12,14,18-21] and coherent anti-Stokes Raman scattering (CARS) [22-25] have enhanced the sensitivity of Raman spectroscopic detection, facilitating rapid imaging. The CARS imaging speed was recently enhanced by widefield [26] and multifocal [27] techniques. However, these nonlinear Raman imaging techniques cannot detect the full spectrum of the sample as well as spontaneous Raman spectroscopy techniques, and the complexity of these systems and the high cost of experimental equipment limit their applicability and use cases. Line-scan Raman spectroscopy is an improvement over single-point detection Raman spectroscopy, replacing a single excitation point with an excitation line, effectively increasing imaging speed by the number of resolved pixels along the line, typically of the order of 100.
The readout speed and noise of the imaging detectors limit the maximum speed. Firstly, the readout noise results in a lower limit for the signal to achieve shot-noise limited detection, and thus a lower limit for the exposure time. Secondly, the ability to expose the detector during readout is required for highest speed. Charge-coupled device (CCD) detectors are widely used in current Raman spectroscopy instruments because of their high quantum efficiency above 80% when back-illuminated, and when cooled to achieve a low dark current and read noise[28,29]. However, due to their few readout taps, often only one, low read noise of 3-5 electrons is only achieved at pixel readout rates below 100 kHz, requiring several seconds to read one frame[30,31]. Furthermore, light typically cannot be collected during readout as the charges are shifted across the imaging array. Electron-multiplying CCDs (EMCCD) can overcome the read-noise limitation, enabling to use faster readout[32-35], but they create excess noise above the photon shot noise. Their readout rate is still limited to a few MHz pixel rate, and their high cost restricts widespread adoption. Recently developed complementary metal-oxide-semiconductor (CMOS) sensors in with each pixel has its own amplifier, and employing one or two taps per row, enable exceptionally fast data readout of hundreds of MHz pixel rates at low readout noise of a few electrons, while also providing a global shutter allowing to collect light while reading the previous frame, and having quantum efficiencies above 50%. Importantly, these sensors are being developed for mobile phones and machine vision and are thus cost-effective and rapidly improving. In 2016, a scientific CMOS camera and a grating spectrometer was used[36], with fast point scanning using about 100mW power, and after subsampling, an effective exposure time of 1.5ms per pixel for cells was reported. In 2024, a custom CMOS sensor was used[37] to construct a miniaturized confocal Raman imaging system for space applications.

In this study, we developed a cost-effective, ultra-fast line scan Raman spectroscopy system using a high-speed CMOS camera, achieving a hyperspectral imaging capability with a spontaneous Raman acquisition rate of 200 fps and 80 kHz spectral rate, superseding conventional line scan Raman spectroscopy methods by two orders of magnitude.
The analysis of the Raman spectra with a low signal to noise in the spectra of single pixels is enabled by the FSC[3] factorisation, which denoises the data by determining from the 1920 spectral points measured for a large number of spatial points ($10^4$-$10^6$) a small number (5-10) of components with their spectra and concentrations. Notably, this is not compressive sensing as used in[38], but is done as post-analysis of the data taken at full spectral and spatial resolution. Rapid Raman imaging on microplastics, tablets, and animal cells is demonstrated, showcasing a wide application potential.

**Results**

**1. High-speed line-scan Raman micro-spectroscopy system**

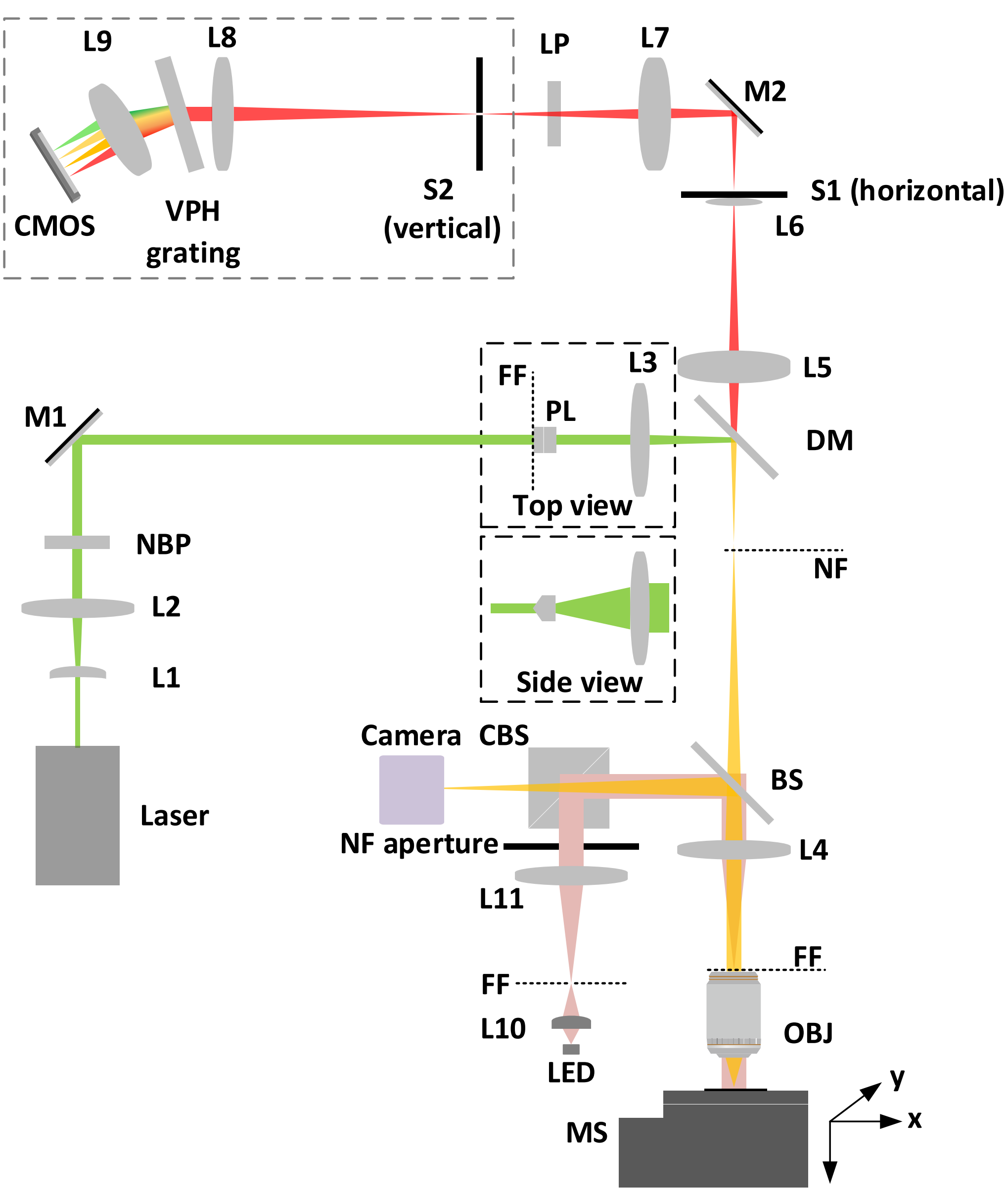

Fig. 1. Sketch of the ultra-fast line scan Raman spectroscopy system using a CMOS camera. L1 to L9 – lenses, S1 and S2 – slits, NBP – narrow band pass filter, M1 and M2 – mirrors, PL – Powell lens, DM –

dichroic mirror, BS – removable beam splitter, CBS – cube beam splitter, OBJ – objective lens, MS – motorized stage, LP – long pass filter, NF – near field image plane, FF – far field image plane, NF aperture – near field aperture, VPH grating – volume phase holographic grating.

Fig. 1 shows the optical design of the ultra-fast line scan Raman micro-spectroscopy system. Apart from the detection, it is similar to our earlier report[30]. Details are given in the Materials and Methods. The line scan Raman technique illuminates the sample with a line rather than a single point and projects an image of the line onto the entrance slit of an imaging spectrometer. We use a 532 nm laser of up to 2 W power which is expanded by lenses L1 and L2, and the filter NBP suppresses background emission at other wavelengths. A Powell lens PL creates a line of beam directions which are focused by lens L3 to vertical line in the near-field (NF) image plane after reflection by a dichroic mirror DM. The tube lens L4 collimates the beam in the horizontal direction towards the infinity corrected objective lens (OBJ), generating the vertical excitation line on the sample. We use a 50× 0.8NA air objective providing a line height at the sample of 105 µm. The sample is mounted on a motorized stage (MS) for x-y plane scanning. For sample positioning and reflection imaging, an illumination by a LED and a camera is used, coupled via a removable plate beam splitter BS.
The signal collection light path implements a dual NF imaging system using lenses L5 and L7. Slit S1 is a horizontal confocal slit with adjustable width, controlling the length of the line passing through. Slit S2 is a vertical confocal slit for line-confocality and input slit of the spectrometer. Long pass filters LP are used to effectively suppress laser transmission.
The spectrometer (gray dashed box in Fig. 1) employs a transmission imaging configuration. It consists of S2, a collimating lens L8, a volume phase holographic (VPH) transmission grating, an imaging camera lens L9, and a machine vision CMOS camera. The CMOS detector is equipped with a global shutter, achieves full frame imaging at 160 fps at a read noise of 2.5 electrons ($e^-$). Its quantum efficiency is around 60% in the relevant range of 540-680 nm. The effective line image on the CMOS sensor is 1.7 mm high, of which we detect the homogeneous central 1.38 mm spanning 400 pixels, equivalent to an effective imaging height of 82.8 µm at the sample. For maximum acquisition speed, we implemented overlapping image acquisition, starting the exposure of a new image while reading out the previous one, as shown in Fig. S1. The overhead time between exposures is only 40 µs, and frame rates of up to 478 Hz for 1920×400 pixels are achievable at 2.05 ms exposure time. To avoid the need for ultrafast positioning, we use a constant velocity motion of the x-y stage across the line illumination, providing a single frame movement across the line given by the velocity divided by the frame rate. For scanning an 82.8 µm × 50 µm area with 4.96 ms exposure time and a stage speed of 100 µm/s, providing 500 nm steps between frames, the total data acquisition time is 0.5 s for 40000 spectra, achieving a speed improvement of several hundred times over previous line-scan Raman imaging including our previous work[30].

## 1. Microplastics

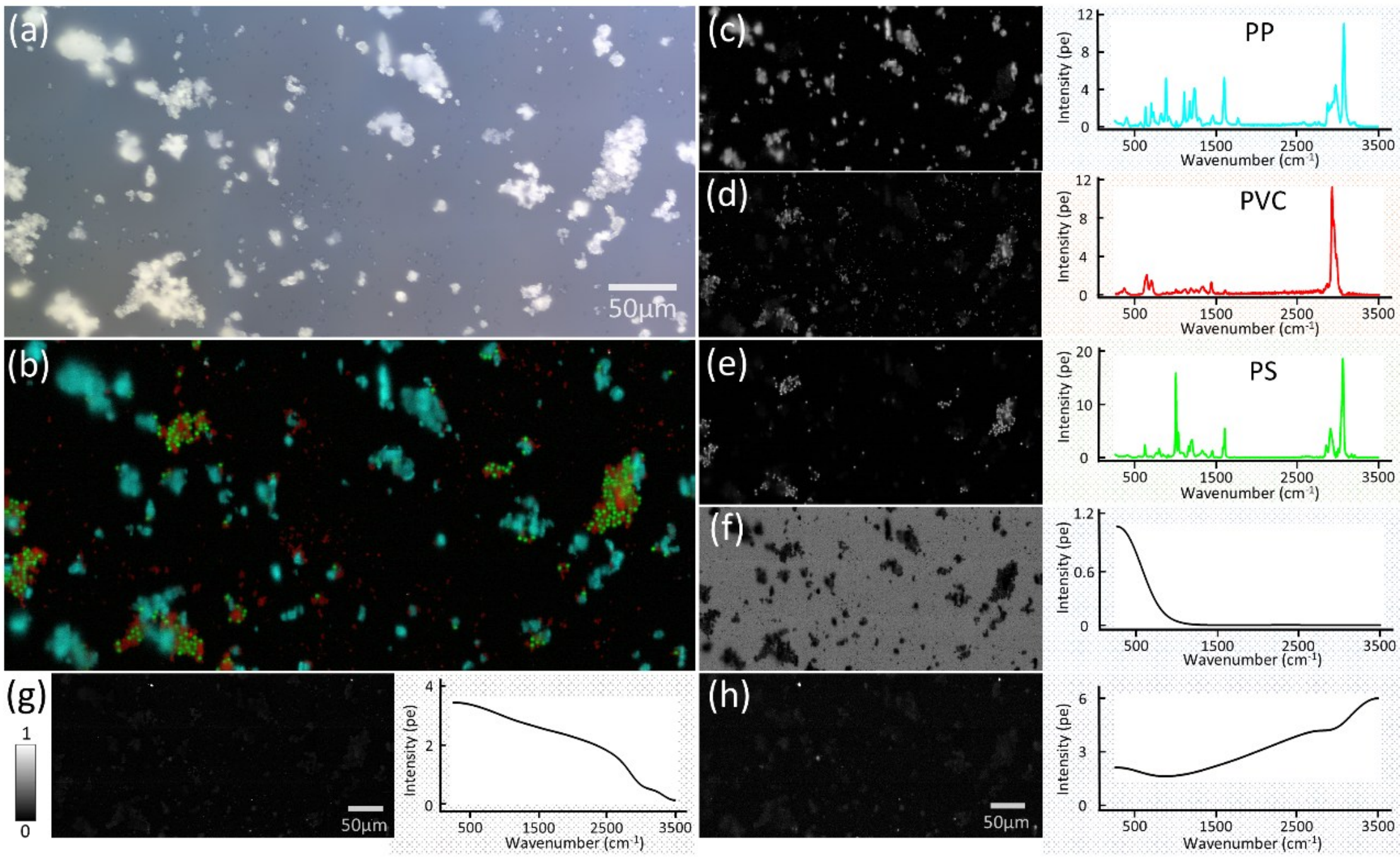

Fig. 2. Large scale (500 µm × 248.4 µm with 1000×1200 pixels) Raman imaging of a microplastic mixture taking 5 ms per line of 400 spectra, 15 s total imaging time. (a) Reflection image (5122 by 10679 pixels). (b) Line scan Raman component concentrations in the region of (a) after $FSC^3$ analysis using six components. Cyan: PP, red: PVC, green: PS. (c-h): concentration images (left) and spectra (right) of individual $FSC^3$ components. The concentrations 0-1 are shown as color value (0-1), with the spectra (in detected photoelectrons per pixel) corresponding to a concentration of 1. (c-e) are Raman components attributed to PP, PVC and PS microplastics, respectively, (f) is the broad Raman component of the fused silica substrate, and (g) and (h) are broad fluorescent components with a spatial distribution indicating contaminations of the microplastics.

Raman imaging results on a microplastic mixture (see Materials and Methods) in a large-area region are shown in Fig. 2. The full-size plots of the spectra are shown in Fig. S11. We performed three scans across the direction of the excitation line, stepped along the excitation line by the detected length of 82.8 µm, resulting in a total imaging height of 248.4 µm. The scanning length was 500 µm, and the stage moved continuously at a constant speed of 100 µm/s. The single frame exposure time was 4.96 ms, resulting in a stepping distance of 500 nm, yielding images of 1000×1200 pixels for a total imaging time of about 15 seconds. The excitation power was 630 mW at the sample. Fig. 2 (a) shows the reflection image of the corresponding region on the substrate, where the microplastics is visible as bright regions of increased scattering.

We used $FSC^3$ to decompose the data into a total of six components, again splitting sharp and broad component spectra, generating concentration images and spectra. (c)-(e) show the components corresponding to PP, PVC, and PS Raman spectra, respectively. The color merged concentration image of these three types of plastics is shown in (b), with red representing PVC, cyan representing PP, and green representing PS, as attributed from their spectra. The Raman images reveal variations in the shapes and sizes of the different particles, enabling clear differentiation even among aggregated plastic materials. PVC particles exhibit varying sizes and shapes. Despite their weak Raman signals, these particles are clearly visible in Raman imaging. The maximum Raman signals per pixel for PP, PVC, and PS were around 11, 11, and 19 photoelectrons (pe), respectively, resulting in spectrally summed signals of 1158, 732, and 1108 pe, respectively. The camera read noise of 2.5 pe is equal to the shot noise of 6.3 pe, thus the Raman signals are detected shot noise limited only for the peak regions. The total number of pixels analysed was 1600, corresponding to a total read noise of about 100 pe, equal to the shot noise of a total signal of 1000 pe. We thus find that for the settings chosen, the detection is at the limit of being shot noise limited. Notably, when reducing the exposure time to 2.05 ms, allowing for the fastest framerate possible with the camera of 478Hz and a spectral rate of 191 kHz, the signal is limited by the read noise, so that the signal to noise scales linear with the exposure time and not with the square root, resulting in a significant deterioration of the results. Notably, this issue could be circumvented by using higher power lasers, or a camera with lower read noise such as scientific CMOS cameras which are now available with a read noise below 1pe.

The components shown in (f-h) have broad spectral features. Component (f) represents the Raman scattering of the amorphous fused silica substrate, which is weak (maximum of 1 pe per pixel, and a sum of 177 pe) and any sharp spectral features have been smoothed in by the FSC[3] analysis treating this component as broad. A fused silica Raman spectrum is quite broad due to the amorphous nature, we compare the substrate spectrum with fused silica. (see Fig. S10). In the concentration image of this component, the areas with particles show a lower concentration due to the scattering by the particles. The stronger Raman spectra are constrained by the data and are accordingly not undergoing smoothing, while the stronger fluorescence components (g,h) seems to be some material which has dried around the microplastics.

The components (g) and (h) represent fluorescence, having peak signals of 3 and 6 pe and summed signals of 3405 and 5475 pe, respectively. They are spatially correlated with all three types of microplastics and show strong spatial variations with a few very localized sources. We interpret these as fluorescent organic molecules present as contaminations in the microplastic. We note that fluorescence cross-sections of molecules can be around 15 orders of magnitude larger than Raman cross-sections[39], so that even very low concentrations of contaminants can give rise to significant fluorescence.

These results demonstrate that the system is capable of high-speed detection and imaging of microplastic particles, with spectral rates up to 80 kHz.

## 2. Cells

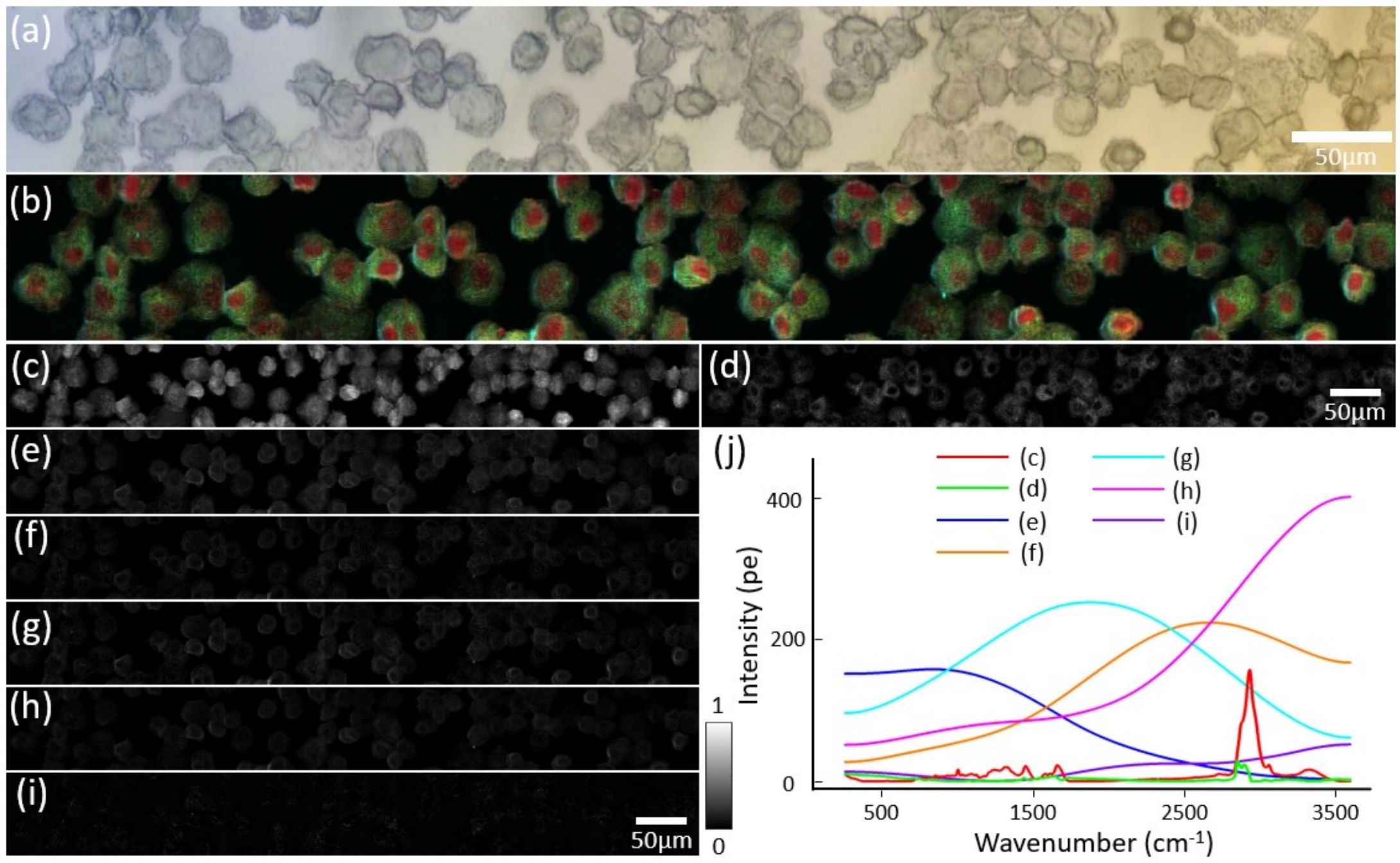

Fig. 3. Line-scan Raman imaging of HepG2 cells on an aluminum-coated glass substrate. (700 μm × 82.8 μm with 1400×400 pixels) taking 200 ms per line of 400 spectra, total imaging time 280 s. (a) Reflection image. (b) Concentration image of the region in (a) after FSC[3] analysis using 7 components. Red and green: Raman components of cells, cyan: sum concentration of fluorescent components. (c-d) Concentration images of the two Raman components. (e-i) Concentration images of fluorescent components. (j) Component spectra.

To demonstrate the system's ability to image individual biological cells, a more difficult task due to the smaller amount of organic material, HepG2 cells dried on aluminium-coated glass or $CaF_2$ substrates were analyzed. The fused silica substrates used for microplastics in Fig. 2 were found to create too much background for these measurements. Enlarged spectra are given in Fig. S12.
The results using an aluminium-coated glass substrate are shown in Fig. 3. The use of aluminium foil as Raman substrate was previously shown[40]. We use a smooth film, suppressing plasmonic hotspots, and we note that the dried cells have a few micrometer thickness, larger than the typical reach of plasmonic hotspots in the 10-50nm range, which are dominating surface enhanced Raman scattering (SERS). We also note that aluminium has a plasmon resonance in the UV, so that at 532 nm the surface plasmon effect is not significant. Thus, the effect of the aluminium substrate is mostly a far-field effect[41]. The collection efficiency is enhanced by the reflection, and the Raman signal is enhanced by the interference of incoming and reflected excitation, providing an average factor of two of excitation intensity. Furthermore, the excitation intensity maxima are also maxima of the Purcell enhancement of the emission, creating a multiplicative effect. The Purcell effect is known for fluorescence lifetime changes near an interface[41]. Combining both effects, the enhancement has the interference pattern over the height $z$ is $(2\sin(z))^4$, having a height average of 6. For the aluminized substrate, the power was limited to 100 mW to avoid photodamage, considering that aluminium is absorbing some 10% of the power, resulting in heating. Notably, this could be improved by using a bulk aluminium slide due to the higher heat conductivity compared to glass, or using aluminized diamond substrates.

Fig. 3 (a) shows the reflection image of cells over the 700 μm × 82.8 μm area imaged by line-scan Raman. The Raman excitation power was 90 mW, and the exposure time was 199.96 ms, with a motion speed of 2.5 μm/s, equivalent to a stepping distance of 500 nm. The imaging time of 280 s for 560000 points corresponds to 0.5 ms per spectrum.

The fixed cells are dried out on the surface and have an oval shape on the substrate. In the $FSC^3$ analysis we found that we required many broad fluorescent components which had spatial shapes reminiscent of thickness profiles and Newton rings, and spectral shapes showing broad oscillations, phase-shifted between the components. We interpret these as Fabry-Pérot fringes occurring due to the reflections at the air-cell and cell-substrate interface. Notably, the cell material of refractive index n=1.45 provides a field reflectivity of $r = (n - 1)/(n + 1) = 0.18$ at the air-cell interface, and thus considering the near unity reflection by the aluminized substrate, a significant spectral intensity modulation between minimum and maximum of about $((1 + r)/(1 - r))^2 = n^2 = 2.1$ is expected. The low thickness of the dried-out cell material in the range of a micrometer suppresses the spatial and directional averaging of this interference. Thus, a single intrinsic fluorescent component acquires a spatio-spectral modulation, requiring many components in $FSC^3$ to represent the detected spectra. The result of $FSC^3$ with 7 components is shown in Fig. 3. Fluorescence is dominating the total signal. We find two Raman components of the cells (shown in (c) and (d)), and five fluorescent components (shown in (e)-(i)). Fig. 3 (b) displays a color merged concentration image of Raman and fluorescence components. The red and green represent the two Raman components and cyan represents the sum of the five fluorescent components. Spectra of all components are shown in (j). The fluorescence components exhibit a spectral modulation expected from Fabry-Pérot fringing. The observed 2000 to 3000 $cm^{-1}$ period corresponds to an optical path length of 5 to 3 μm, and thus a thickness of a dried cell layer of 1.7 to 1 μm assuming a refractive index of 1.5, consistent with the dried cell thickness. The different spectra show relative phase shifts of the oscillations to accommodate the different thicknesses across the sample. The Raman component (d) in green mainly consists of lipids within the cells, particularly concentrated in the cytoplasmic regions (membranes of endoplasmic reticulum, Golgi apparatus, mitochondria). The Raman component (c) in red includes a wider range of cellular substances such as nucleic acids, proteins, and residual water, is primarily visible in the remainder of the cell nucleus. These qualitative assignments to materials is done using previous literature [Modern Vibrational Spectroscopy and Micro-Spectroscopy Max Diem, 2015, ISBN 1118824865]. The fluorescence component uniformly distributes across various cellular regions, exhibiting a more pronounced presence in certain regions of the cell membrane.

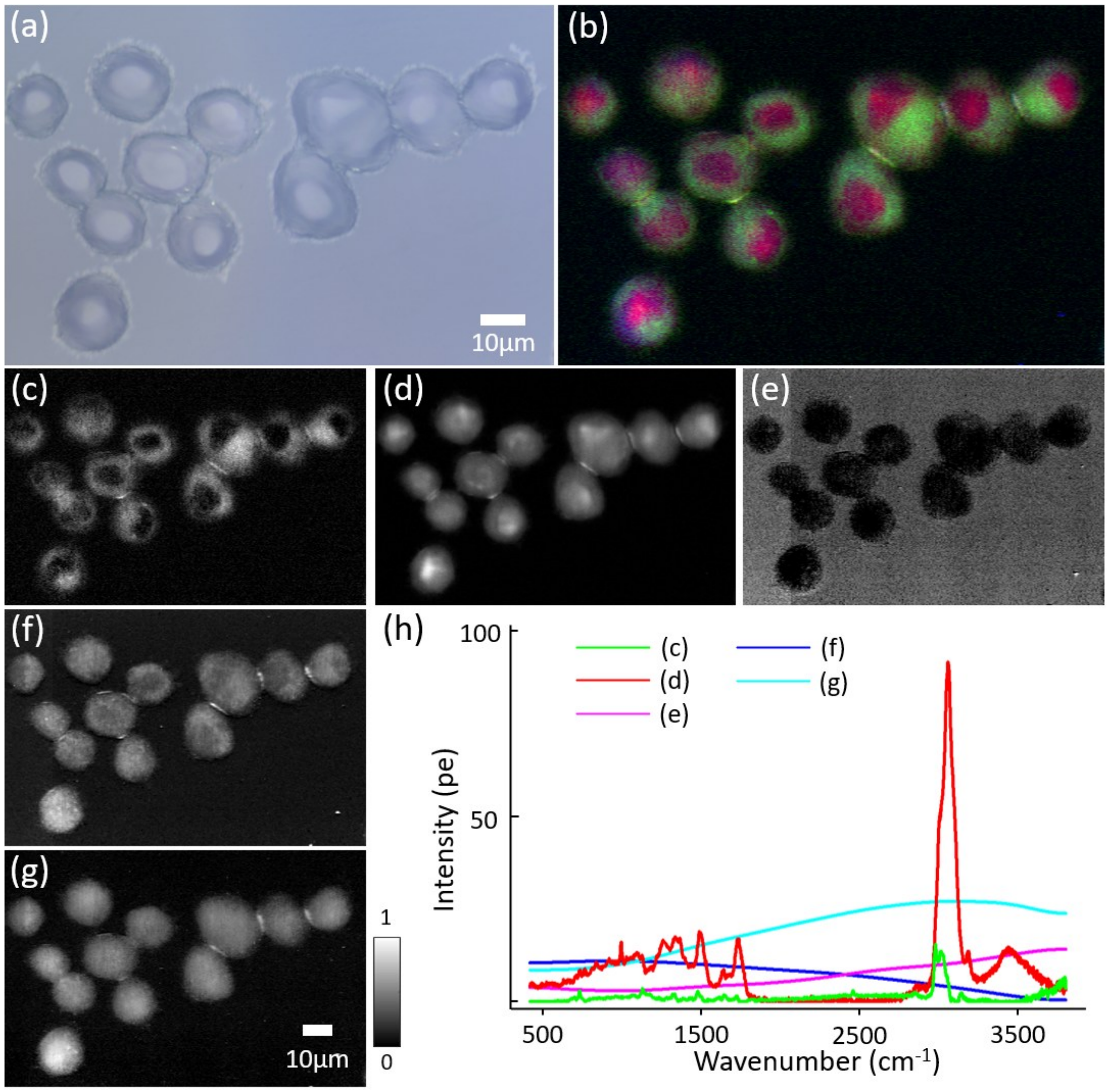

Fig. 4. Line-scan Raman imaging of HepG2 cells on a $CaF_2$ substrate (120 µm × 82.8 µm with 240×400 pixels) taking 200 ms per line of 400 spectra, total imaging time 48 s. (a) Reflection image. (b) Line-scan Raman component concentration image of the region in (a) after FSC$^3$ analysis. Red and green: Raman components of cells, blue: fluorescent components of cells. (c) and (d) concentration images of two Raman components. (e) concentration image of substrate component. (f) and (g) concentration images of two fluorescent components of cells. (h) component spectra.

Using a $CaF_2$ substrate of refractive index 1.43, there is virtually no reflection at the interface to the cells, and thus the interference fringes in the fluorescence spectra are avoided. The corresponding imaging results are shown in Fig. 4 on an 82.8 µm × 120 µm area with 240×400 pixels, and 199.96 ms exposure time, 0.5 ms per spectrum, total imaging time 48 s. The full size plots of the spectra are shown in Fig. S13. The reflection image (a) shows the cells investigated. Using FSC$^3$ we again found two Raman components shown in (c) and (d). However, we now have one substrate fluorescence component (e) and only two cellular fluorescent components (f) and (g), which resemble the distribution of the Raman contributions, without Fabry-Pérot fringing. The peak signals for the components (c-g) are 16, 92, 14, 11, and 27 pe with sums of 2257, 14777, 12400, 11704, and 34083 pe, respectively. The excitation power was 530 mW, which is 6 times higher than used on the aluminum substrate, but the Raman components are still about a factor 1.5 times weaker, with the other acquisition settings the same. This is consistent with the expected enhancement of the Raman signal by the aluminized substrate discussed above. (b) shows the color merged concentration image of Raman and fluorescent components. The Raman components of (c) and (d) are respectively marked in green and red, and the sum of the fluorescence components in (f) and (g) are both denoted in blue. The concentration images and spectra of the Raman components are similar to the results on the aluminized substrate, Fig. 3. The green corresponds to the cytoplasmic area, predominantly consisting of lipid membranes, and the red encompasses a broader range of cellular materials, including nucleic acids, proteins, lipids, and water. Meanwhile, the blue fluorescence components are uniformly distributed within the cells.
Compared to the microplastics data in Fig. 2, the relatively stronger background fluorescence requires a larger overall signal to separate the Raman components due to the increased shot-noise. The average fluorescence of about 50 pe creates a shot noise of 7 pe, triple the read noise of the camera, requiring correspondingly at least three times the signal, not considering systematic influences.

### 3. Pharmaceutical tablets

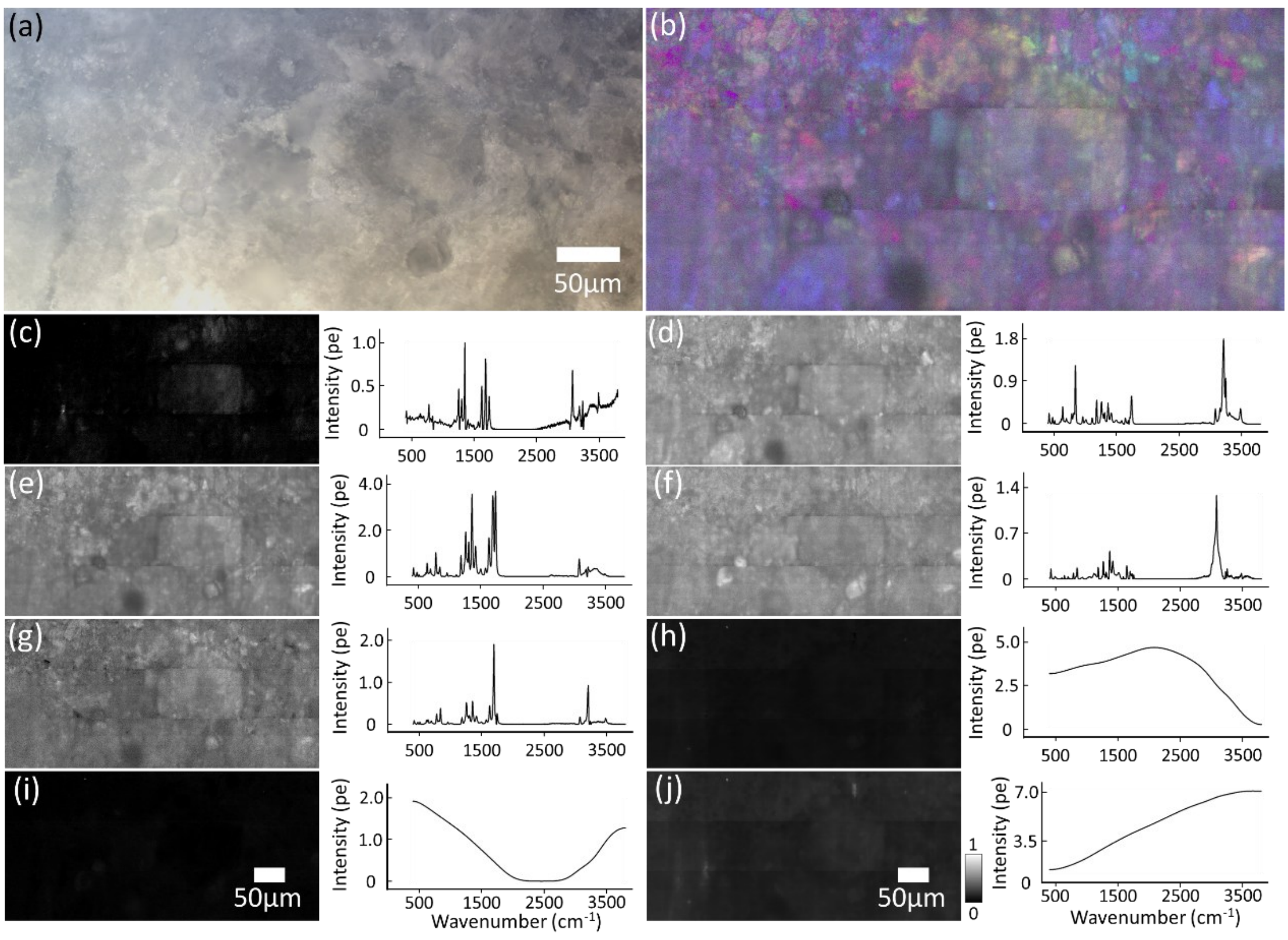

Fig. 5. Line-scan Raman imaging of an acetaminophen tablet (500 µm × 248.4 µm with 1000×1200 pixels) taken at 5 ms per line of 400 spectra. (a) Reflection image. (b) Raman component concentration image of the region in (a) after 8 component FSC[3] analysis showing the components (e) in red, (g) in green, and (d) in blue. Concentration images and spectra of Raman (c-g) and fluorescence (h-j) components.

The imaging results on the acetaminophen tablets are shown in Fig. 5. The full size plots of the spectra are shown in Fig. S14. Panel (a) shows the reflection image of the scanned region of 248.4 µm height and 500 µm length. The excitation power was 630 mW, and the exposure time was 4.96 ms, with a velocity of 100 µm/s, equivalent to a stepping distance of 500 nm, as for the microplastic data Fig. 2. The total scanning time for this area was 15 s. Before analysis, the image data underwent spectral tilt and distortion correction to eliminate the small differences of the spectral calibration for different rows within 1-2 pixels, caused by small alignment errors and imaging distortions (see Fig. S3). This was required here as the Raman spectra of the organic crystals have narrow linewidths of 10-20 $cm^{-1}$, close to the resolution of the instrument. Five Raman components and three fluorescence components were identified by FSC[3]. The Raman components shown in (c,d,e,g) feature similar characteristic peaks of acetaminophen compounds, but with clear differences in concentration distribution. (b) shows the color merged concentration image integrating the components (d,e,g) as RGB. These differences in spectra may be caused by changes in the crystal orientation of acetaminophen. The chemical composition of component (f) has not been identified. The morphology and orientation of crystals play a crucial role in the bioavailability and stability of drugs. Despite the chemical consistency of acetaminophen, minor variations in crystal morphology and orientation can significantly impact the drug's performance. This study highlights the potential of this technique for rapidly identifying different crystalline forms of the same compound in tablets.

Importantly, the tablet is a diffuse scatterer, so that sub-surface signals are significantly influencing the collected signal, as can be seen by the partially smooth shapes in the concentrations. The line-scan Raman instrumentation presented here thus lends itself to ultrafast spatially offset Raman measurements to investigate the depth distribution within diffuse scatterers[42].

**Discussion**

We have developed an ultra-fast line-scan Raman imaging technique based on a cost-effective machine-vision CMOS detector with high readout rates and low read noise, and a simple lens-based holographic transmission grating imaging spectrometer. At the same time as reducing the cost significantly, we have shown a step change in Raman imaging speed, accelerating the imaging by one to two orders of magnitude compared to previous line-scan Raman imaging techniques, and by three to four orders of magnitude compared to standard point confocal Raman. Through the continuous acquisition mode with continuous sample motion, imaging at 80 kHz spectral rate was demonstrated, allowing to acquire a megapixel Raman image within 12 seconds. Such increased speed is enabled by the lower read-noise of the CMOS detector compared to CCD detectors at the required high readout rates.

The fast scanning reduces photodamage by heating compared to slower instruments. No photodamage was observed for the powers and scan speeds used. For the aluminized substrate, the power was limited to 100 mW, as aluminium is absorbing some 10% of the power, resulting in heating. Notably, this limit could be increased using a bulk aluminium slide or aluminized diamond substrates due to the higher heat conductivity compared to glass.

Due to the very high speeds of our acquisition, a real time display of the raw data (up to 300MB/s) is beyond what humans can visually process. The image processing is done presently off-line as post processing. However, using a mainstream modern GPUs in the data acquisition PC (such as an Nvidia RTX 5070) it is possible to provide real time factorization into the $FSC^3$ components which could then be shown as concentration maps in real time. We note that 478 frames per second and 191 kHz spectral rate are supported by the hardware if Raman signal strength allows. Furthermore, using recent advances in CMOS detector technology reducing the read noise to below 1 electron (for example the Hamamatsu C15550-22UP), the speed for shot-noise limited detection could be increased by an order of magnitude. We introduce an improved $FSC^3$ analysis method doing away with fluorescence fitting and instead retrieving both fluorescence and Raman components, keeping both types of signals and avoiding fitting artefacts. This allows us to rapidly identify the spatial distribution of multiple Raman and fluorescence components within samples. In terms of applications, this technology demonstrates efficient detection capabilities for a variety of samples, including microplastics, biological cells, and tablets. The low cost, portability, and high performance indicate great advantages for commercialization and uptake, paving the way to its widespread application in research, environmental monitoring, and industry.

## Materials and Methods

### 1. High-speed line-scan Raman micro-spectroscopy system

We use a 532 nm laser (Laser Quantum, UK, GEM532) with $TEM_{00}$ mode and 2 W maximum output power and employ lenses L1 (Thorlabs, USA, ACN127-025-A, f=-25 mm) and L2 (Thorlabs, USA, AC254-050-A, f=50 mm) to expand and collimate the beam to 1.8 mm size at $1/e^2$ intensity. A bandpass filter NBP (Semrock, USA, LL01-532-12.5) of 2 nm width centered at 532 nm is used to suppress background emission at other wavelengths. An achromatic lens L3 (Thorlabs, USA, AC254-060-A, f=60 mm) focuses the line of beam directions created by the Powell lens PL (Laserline Optics, Canada, LOCP-8.9R05-1.8, fan angle 5 degrees, input beam diameter 1.8 mm) placed in a far-field (FF) plane of the objective to a 5.2 mm high vertical line in the near-field (NF) image plane after reflection by a dichroic mirror DM (Chroma, USA, RT532rdc). The tube lens L4 (Thorlabs, USA, AC254-200-A, f=200 mm) collimates the beam in the horizontal direction towards the infinity corrected objective lens (OBJ), generating the vertical excitation line on the sample. We use a 50× 0.8NA air objective (TU Plan Fluor, Nikon, Japan) with a 6.4 mm back focal plane (BFP) diameter, so that the 6 mm beam size at the BFP provides a fill factor of 0.94, and the line height at the sample is 105 µm. The sample is mounted on a motorized stage (MS) (Jiancheng Optoelectronics, China, X-Y-Z-V05, travel 50 mm, minimal incremental motion 50 nm, maximum speed 3 mm/s) for x-y plane scanning. For sample positioning and reflection imaging, an illumination by a LED module is coupled to the objective via a cube beam splitter CBS (Thorlabs, USA, CCM1-BS013, 50:50 R:T) and a removable plate beam splitter BS (Thorlabs, USA, BSX10R, 90:10 R:T), and the reflected sample image is detected via a camera (Dothinkey, China, M3ST502(M)-H, 3072×2048 pixels). The LED module is composed of a white light LED (Thorlabs, USA, MNWHD2, 4900K), an aspheric condenser lens (Thorlabs, USA, ACL25416U-A, f=16 mm), and an achromatic lens (Thorlabs, USA, AC254-100-A, f=100 mm), and an iris aperture in the NF plane to control the diameter of the illuminated area on the sample.

The signal collection light path implements a dual NF imaging system using two Steinheil achromatic triplet lenses, L5 and L7 (Edmund, USA, #67-422, f=50 mm), optimized for 1:1 imaging. Slit S1 (Thorlabs, USA, VA100C) is a horizontal confocal slit with adjustable width, controlling the length of the line passing through. Slit S2 (Princeton Instruments, USA) is a vertical confocal slit, ensuring line-confocality and controlling spectral resolution. A biconvex lens L6 (Thorlabs, USA, LB1014-A, f=25 mm) is mounted on S1 to image the far-field (FF) onto the grating of the spectrometer, avoiding vignetting along the line. Two long pass filters LP (Semrock, USA, BLP01-532R-25) with a combined optical density of at least 12 at 532 nm are used to effectively suppress laser transmission. The filters transmit more than 50% above 543 nm, corresponding to Raman scattering with wavenumbers above 380 $cm^{-1}$, and more than 86% above 500 $cm^{-1}$.
The spectrometer (gray dashed box in Fig. 1) employs a transmission imaging configuration. It consists of S2, a collimating lens (L8), a volume phase holographic (VPH) transmission grating, an imaging camera lens (L9), and a machine vision CMOS camera. The light passing through Slit S2 is collimated by L8 (Thorlabs, USA, AC254-150-A, f=150 mm) and enters the VPH grating (Wasatch Photonics, USA, WP-1000/600-50.8). The VPH grating features a line density of 1000 lines/mm, a central wavelength of 600 nm, and an incident angle of 17.5°, providing a first-order average diffraction efficiency of about 90% for both p and s polarization within the 540-680 nm range. The camera lens L9 (Basler, Germany, C23-5028-5M-P, f=50 mm, F2.8-F16) focuses the dispersed light onto the CMOS camera (Basler, Germany, a2A1920-160umBAS, 1920×1200 pixels using a Sony IMX392 sensor chip), creating a 3:1 conjugate image of Slit S2. The CMOS detector is equipped with a global shutter, achieves 8-bit full frame imaging at a rate of 160 fps at a read noise of 2.5 electrons ($e^-$). Its quantum efficiency is around 60% in the relevant range of 540-680 nm. The size of each pixel on the CMOS sensor is 3.45 μm square, translating to a corresponding pixel size of 10.35 μm square at Slit S2. The grating provides a dispersion of about 2 $cm^{-1}$ per pixel, thus a total range of about 3800 $cm^{-1}$, suited to cover the relevant range of vibrations occurring in materials including water, with a spectral resolution (FWHM) of about 5 $cm^{-1}$. The objective lens provides a magnification of 50 on the image plane of Slit S2, so that the pixel size is 207 nm at the sample. The slit S2 was typically set to 50 μm, corresponding to 1 μm on the sample. This is 2 times larger than the excitation line width and ensures that most of the Raman emission is collected. The dispersion is about 2 $cm^{-1}$ per pixel on the camera. Considering the 3.45 μm pixel size and the 1:3 demagnification from slit to camera, a 50 μm slit width corresponds to about 10 $cm^{-1}$ resolution. The wavenumber range is 256-4048 $cm^{-1}$, as can be seen in Fig.S8.
In our measurements, we use the maximum opening F2.8 of L9, providing an input aperture of 18mm diameter. This accommodates the beam entering the grating of 4.8 mm diameter as well as the dispersion by the grating, avoiding vignetting. The effective line image on the CMOS sensor is 1.7 mm high, of which we detect the homogeneous central 1.38 mm spanning 400 pixels, equivalent to an effective imaging height of 82.8 μm at the sample. We used 30 dB analog gain of the sensor, providing 1.34 e per count at 8bit readout, sufficient to be dominated by the sensor read noise of 2.5 e. A black level of 30 is used, corresponding to 2 counts at 8 bit readout, to avoid zero clipping.
Notably, many Raman systems in literature use spectrometers with a small numerical aperture at the exit, for example a Czerny-Turner spectrometer[36] of F/6.5 opening, 2.5 times smaller than the F/2.8 used here, thus the 3.45 μm pixels we use gather the same input entendue as 8 μm pixels at F/6.5, larger than the 6.5um pixels of the sCMOS camera used in[36]. Smaller pixels have lower dark currents,, allowing to use an uncooled CMOS camera of much lower cost (£500) than the cooled sCMOS camera (£10000) used in[36].
The total optical transmission from sample to detector is 51%, determined by the following individual transmissions of the optical elements: 80% Objective, 86% filters, 90% camera lens, 5 lenses 99% each, the dichroic mirror (DM) 96%, and the VPH (90%). Considering the 60% QE of the camera results in 34% Raman photon detection efficiency.
The 0.8NA objective collects 40% of the solid angle for the aluminized reflective substrate using a 2pi total solid angle. For the transparent substrates, the effective solid angle in the substrate is 2pi times the refractive index of 1.52 for glass slides and 1.43 for $CaF_2$, resulting in a 12% and 13% collection by the objective, respectively.
In our previous work using a spectrometer and a CCD detector[30], a power of 100-200 mW was used at the sample, limited by the damage at the slower scan speed used. The detector QE was higher (90%), the transmission of the spectrometer was lower, so that the detection efficiency in total was similar. However, the read noise of the CCD at 2MHz pixel rate was 11pe, and thus needs 21 times more signal than the present camera to be shot-noise limited.
As we have discussed in the manuscript, the spectrally integrated shot noise of the signal and the spectrally integrated read noise are similar for the data of Fig.2. Therefore, as in Fig.6 of our previous work[30], the detection is close to the shot noise limit. Notably, our previous work used a 0.6NA objective, which collects half the solid angle of the 0.8NA objective used here.

**2. Data Preprocessing**

The Raman data collected by the CMOS camera is a two-dimensional array 1920 × 400 pixels. The data of each row of 1920 pixels provides a Raman spectrum, with the spatial position along the excitation line imaged on the 400 pixels high columns.
The raw data undergoes a series of preprocessing steps, including offset and dark subtraction, hot pixels correction, sensitivity correction and denoising, similar to our previous work[30]. Each processing step is detailed in the supplement.
Firstly, the dark signal is subtracted using an average of hundred images taken with the Raman excitation laser blocked but otherwise equal settings. Notably, the machine vision camera used is working at room temperature and has many so-called “hot” pixels which have a significantly larger dark current. The approach for correcting hot pixels is illustrated in Fig. S2. The hot pixels are identified using an average image of ten frames with 5s exposure time in a dark environment and calculate a threshold by summing the mode and three times the standard deviation of the intensity values in this average image. Pixels with values exceeding the threshold were categorized as hot pixels. The values of hot pixels are set to the average value of adjacent pixels in the spatial (column) direction, thus keeping the spectral position and interpolating over the spatial position only. To correct the spatial and spectral response to a spatially homogeneous and spectral sensitivity calibrated data, a SRM 2242a (National Institute of Standards & Technology) standard material[43] is used. The correction for each pixel is obtained using the ratio between the value measured on this material and the material standard curve, normalized across all pixels to unity average. The data pixel values are divided by this correction. Finally, singular value decomposition (SVD) is used to denoise the data[44], retaining only singular vectors showing physically meaningful spatial and spectral shapes above noise, as detailed in the supplement . The large number (typically $10^6$) of spectra in the acquired data, many times the number of spectral points, makes the denoising by SVD reliable.

**3. Data analysis**

After preprocessing, the spectra were analyzed using Factorisation into Spectra and Concentrations of Chemical Components ($FSC^3$).[45-48] Firstly, the standard $FSC^3$ algorithm was used, which employs a non-negative matrix factorization (NMF) algorithm. The hyperspectral data are decomposed into a linear combination of chemical components defined by their non-negative Raman spectra and spatial v/v concentration distributions. The number of spectra is manually chosen to retrieve all relevant chemical components including the surrounding matrix, while avoiding splitting of components which develops when using to many. The component spectra are normalized to yield maximum concentrations of one across the spatial maps. The $FSC^3$ method used is separating fluorescence and Raman components using their spectral structure. To separate the components in either spectrally smooth or sharp components, the iteration contains additional procedures either smoothing the spectra which are already smooth, or subtracting from sharper spectra a fraction of their mean. This results in the iterative transfer of broad offsets in Raman components to the broad fluorescence components. There are a range of parameters and details in this fluorescence / Raman separation in addition to the $FSC^3$ method which will be published in detail elsewhere.

The sum of the component spectra weighted with the concentration provides the photoelectron counts in the raw spectra, and the concentrations have a maximum of 1 in the data.

Our previous work used a robust fit prior to $FSC^3$ to remove fluorescence background based on its spectrally broad features, a method which is widely used in the field, albeit with a variety of algorithms and corresponding spread in results. Apart from the underlying issue that also Raman spectra can be broad and thus are also removed by the background fit to varying degrees depending on the algorithm and the parameters chosen, the background fit is done on individual spectra and thus is prone to create spatial artifacts. In the present work we have eliminated the background fit and instead use $FSC^3$ to separate components which have spectra resembling Raman or fluorescence. This is done by adding to the $FSC^3$ iteration loop a modification of spectra, nudging them towards being either spectrally smooth or having spectral regions close to zero intensity, depending on their smoothness. Since $FSC^3$ uses all spectra available and creates a common spectral basis, this approach removes the spatial noise of the background fit and is often able to retain realistic Raman spectra including their spectrally broad components, as well as retrieving fluorescence components and their spatial distribution. Details of this method will be published elsewhere.

**4. Sample preparation**

**4.1 Microplastics**

To demonstrate the high-speed detection capabilities of the system, we used three types of microplastics as sample materials: polyvinyl chloride (PVC) and polypropylene (PP) in the form of microplastics powders (ShuangFu, China) with irregular shapes, and polystyrene (PS) in the form of nominally spherical beads (YUAN Biotech, China, YB2003) with 3 μm size.

For the powders, we prepared suspensions by mixing 0.1 g of each with 10 mL of ultrapure water and 10 mL of absolute ethanol. The concentration of the stock PS bead suspension was 25 mg/ml, and 0.5 ml of it was added to 9.5 ml of a 1:1 v/v water/ethanol solution. Each suspension was ultrasonicated for 30 minutes to disperse the particles. Then, the three suspensions were mixed in equal amounts to prepare a mixture suspension, and 2 µL of the mixture suspension were pipetted onto a fused silica substrate (75 × 25 × 1 $mm^3$) forming a drop of about 5 mm diameter, and subsequently air-dried.

### 4.2 Biological cells

HepG2 cells were selected for our measurements. These human hepatocellular carcinoma cells are of adherent type. Cultivation was carried out in DMEM medium enriched with 10% fetal bovine serum, under conditions of 37°C and 5% $CO_2$. After 14 hours of culture, the cells were lifted off using a 0.25% trypsin solution. Subsequently, the cells were fixed with 4% methanol buffer and washed with phosphate-buffered saline (PBS). Following washing, cells were spotted onto aluminized glass or calcium fluoride ($CaF_2$) substrates with a pipette, 2-3 µL of sample was added dropwise to each substrate, and then air-dried under ambient conditions, leaving a covered area of about 5 mm diameter on the substrate.

### 4.3 Pharmaceutical tablets

Pressed tablets were used to investigate the distribution of the contained substances, including variations in crystalline compounds. Specifically, tablets (Liao Yuan City Baikang Pharmaceutical Co., Ltd., China) containing as active ingredient 0.5 g of acetaminophen were used. In addition to the active ingredient, the tablets are also known to contain a variety of excipients, including starch, thiourea, magnesium stearate, sodium carboxymethyl starch, and dextrin. The full information of tablets given by the supplier (see Table.S1). The tablets were cylindrical in shape, with a diameter of 10 mm and a thickness of 3 mm (see picture in Fig. S5). To ensure the stability of the scanning process, double-sided tape was used to adhere the tablets to the slides, and clamps were used to keep the slides in place on the displacement stage. We note that inhomogeneities in the surface of the tablets themselves can still lead to focusing differences.

## Author Contributions

Qingyi Wu: Investigation, Formal analysis, Data curation, Writing – original draft, Writing – review & editing. Xusheng Tang: Investigation, Visualization, Data curation, Writing – original draft, Writing – review & editing. Francesco Masia: Software, Formal analysis. Peng Liang: Investigation. Hao Peng: Data curation. Lindong Shang: Visualization. Yuntong Wang: Investigation. Yue Qu: Visualization. Wolfgang Langbein: Conceptualization, Methodology, Software, Writing – review & editing. Bei Li: Conceptualization, Methodology, Funding acquisition.

## Conflict of interest

The authors declare no competing interests.

## Data availability

All raw data generated in this study are available from the corresponding author upon reasonable request.

**Acknowledgments**

Funded by National Key Research and Development Program of China. Grant No. 2024YFF0726100.

Funded by National Natural Science Foundation of China. Grant No. 62304222.

Funded by Chinese Academy of Sciences President's International Fellowship Initiative. Grant No. 2022VBA0028.

# Ultra-high-speed line-scan Raman imaging – Supplementary Information

**Qingyi Wu[1,2,3,#], Xusheng Tang[1,2,3,#], Francesco Masia[5], Peng Liang[1,2,3,4], Hao Peng[1,2,3], Lindong Shang[1,2,3], Yuntong Wang[1,2,3], Yue Qu[4], Wolfgang Langbein[6,*] and Bei Li[1,2,3,*]**

[1]State Key Laboratory of Advanced Manufacturing for Optical Systems, Changchun Institute of Optics, Fine Mechanics and Physics, Chinese Academy of Sciences, Changchun 130033, P. R. China

[2]University of Chinese Academy of Sciences, Beijing 100049, P. R. China

[3]State Key Laboratory of Advanced Manufacturing for Optical Systems, Changchun 130033, P. R. China

[4]Haining High-tech Research Institute, Jiaxing, Zhejiang, 314408, PR China

[5]School of Biosciences, Cardiff University, Cardiff, CF10 3AX, UK

[6]School of Physics and Astronomy, Cardiff University, Cardiff, CF24 3AA, UK

*** Correspondence:**
Wolfgang Langbein and Bei Li
langbeinww@cardiff.ac.uk and beili@ciomp.ac.cn

[#] **These authors contributed equally:** Qingyi Wu, Xusheng Tang. Qingyi Wu and Xusheng Tang are the co-first authors.

## 1. Data acquisition

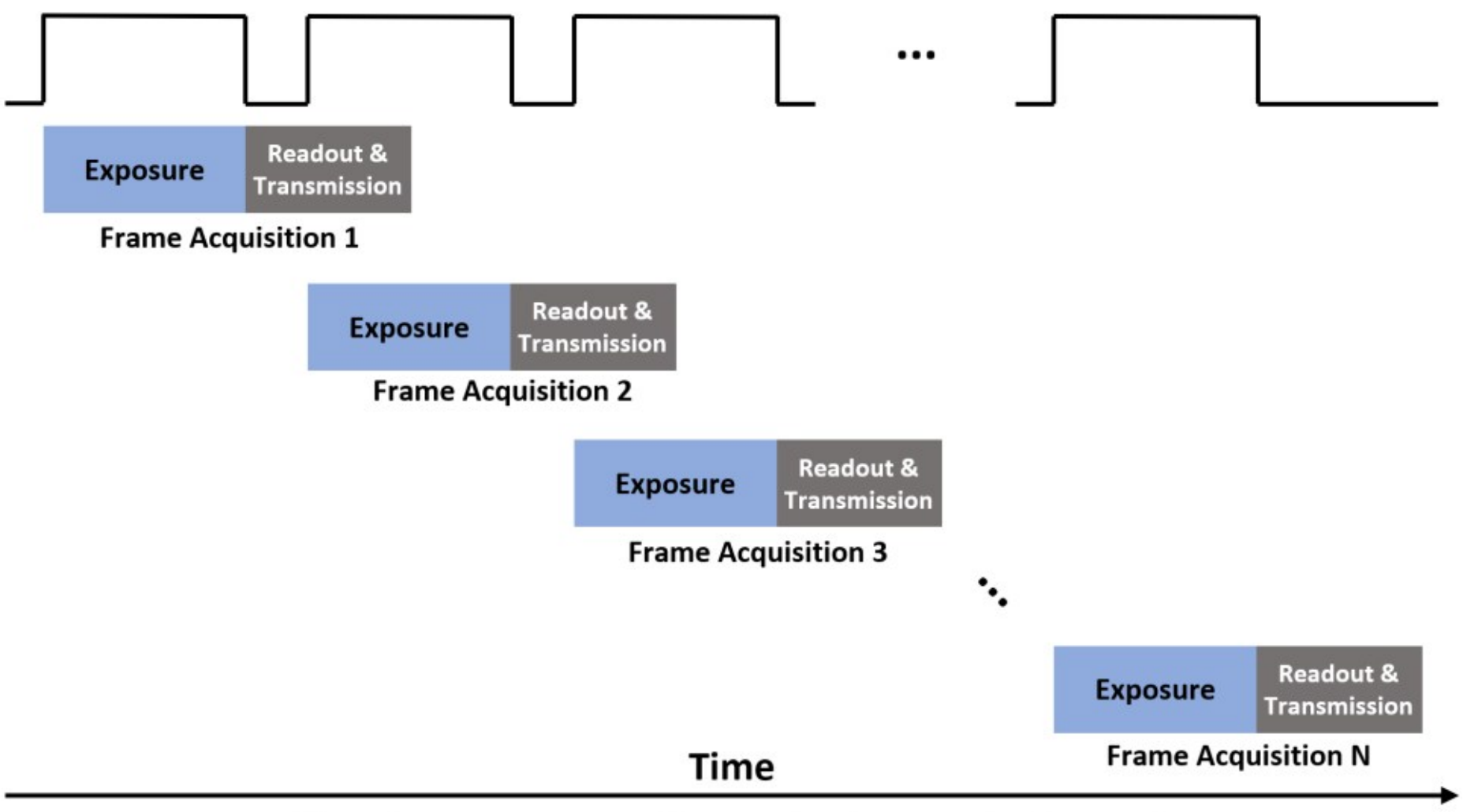

Fig.S1. Diagram of acquisition timing

Fig. S1. illustrates the acquisition timing for the CMOS camera. The CMOS camera is equipped with a global shutter. To achieve higher frame rates, the camera employs overlapping acquisition, which involves exposing a new image while the sensor data from the previous image is being read out. This operational mode significantly reduces the impact of the camera's data readout and transmission times on the overall detection time. Simultaneously, the x-y stage moves continuously at a constant speed given by the camera frame rate times the distance per frame. The time between exposures is as low as 40 µs for the Sony IMX392 chip used.

## 2. Hot pixel correction

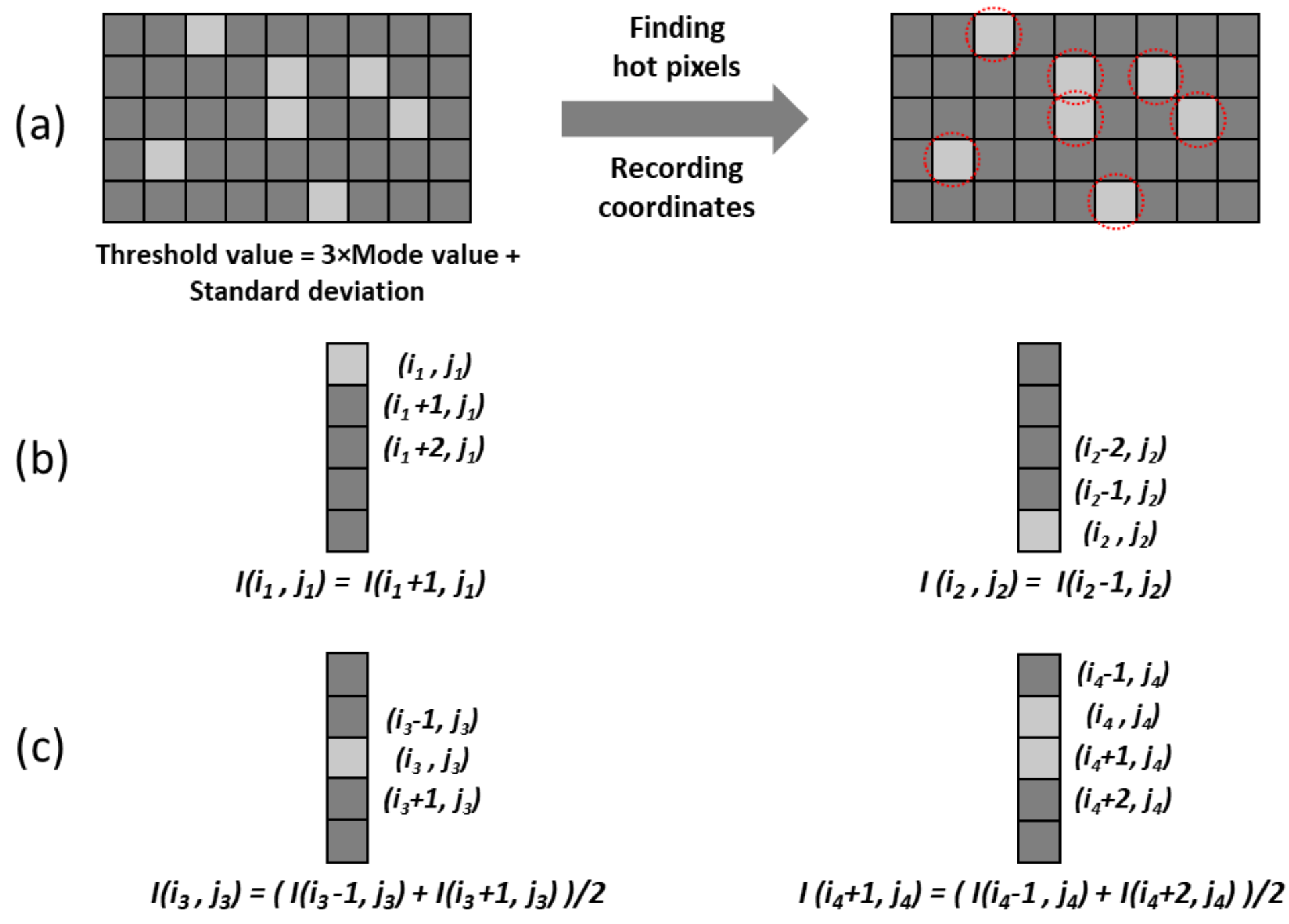

Fig. S2. Sketch of hot pixels correction method

Fig. S2 shows the method used for hot pixels correction. An average of ten frames with 5 s exposure time in a dark environment is taken and a threshold is determined by summing the mode of the distribution and three times the standard deviation of the intensity values in this average image. Pixels with intensities exceeding the threshold are categorized as hot pixels, as shown in Fig. S2 (a). Approximately 0.5% of the overall pixel count was comprised of hot pixels defined in this way. To avoid compromising spectral position information, we corrected the intensity of hot pixels by using the average intensity of adjacent pixels in the column direction. If the hot pixel is in the top or bottom row as shown in Fig. S2 (b), and we use the average intensity of the two adjacent pixels below or above for correction. Otherwise, we use the average intensity of the two adjacent pixels above and below to replace the hot pixel's intensity, as shown in Fig. S2 (c). We note that for typical exposure times of 5 to 200 ms used in the data shown in the manuscript, a reduced number of hot pixels could be used.

### 3. Tilt and Distortion Correction

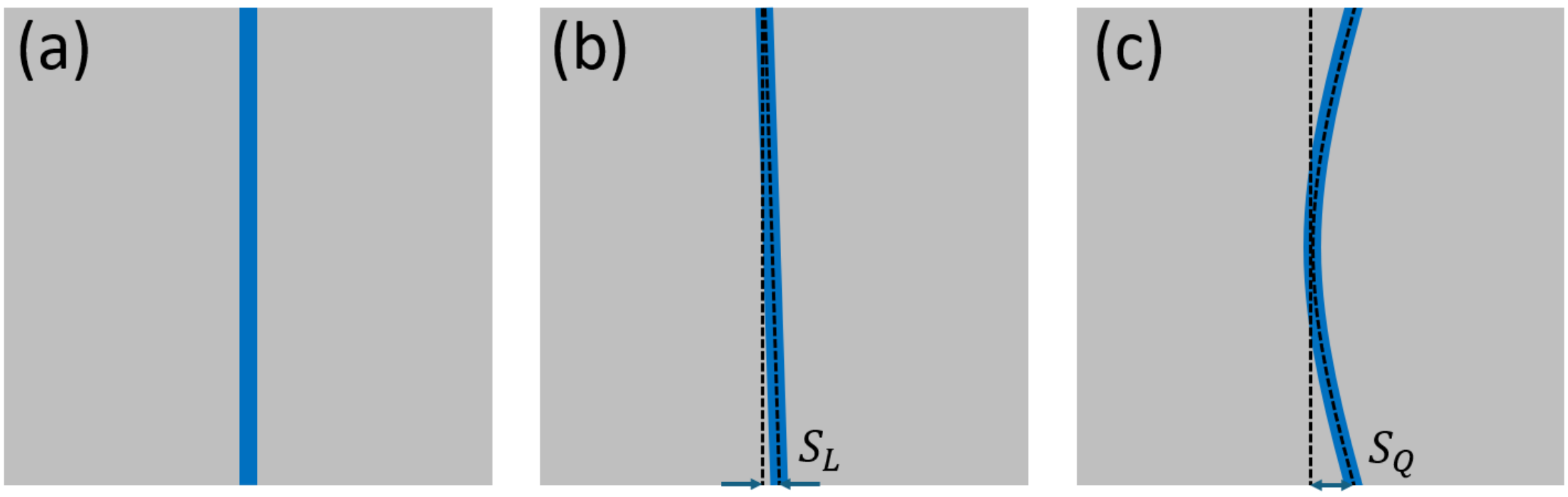

Fig. S3. Schematic diagram of the tilt and distortion of the line. (a) shows a vertical line spot on the CMOS sensor, (b) shows the tilt of the line, and (c) shows a quadratic cushion distortion of the line, with the top and bottom bent in the same direction.

The pixels in the camera are identified by their coordinates $(s, y)$, where $s$ represent the horizontal (spectral) direction with $w$=1920 pixels, and $y$ is the position along the vertical (position) direction with $h$=400 pixels. We are correcting here only the spectral direction, to ensure a spectral calibration independent of the vertical position. Because of residual camera tilt and the distortion of the imaging by the camera objective, the intensity expected at pixel $(s, y)$ is located at pixel $(s', y')$ with $s'$ given by

$$s' = a + b(s - s_0) + c(s - s_0)^2 + d(y - y_0) + e(y - y_0)^2 + f(s - s_0)(y - y_0) + g(s - s_0)(y - y_0)^2.$$

The intensity at $(s', y)$ is mapped back onto $(s, y)$ using linear interpolation of each row $(y)$ independently. We use $a = s_0 = 960.5$, $y_0 = 200.5$ given by the center of the spectrum, and start wit $c = d = e = f = g = 0$, and $b = 1$, providing no correction. A relative tilt of the line around the center of the chip as shown in Fig. S3 (b) can be corrected using the parameter $d$, calculated using the shift observed between the top and bottom of the camera $S_L$, as $d = S_L/h$. A quadratic cushion correction centered at the center of the camera as shown in Fig. S3 (c) can be corrected using the parameter $e$ calculated from the shift of the line at the top of the camera with respect to the position at the center $S_Q$, $e = S_Q/(h/2)^2$.

The results of the calibration of the spectral data using this method are shown in Fig S4, using tablets as samples. In the 1920×400 resolution image shown in Fig. S4 (a), we have selected the spectral data in rows 1, 200, and 400 for comparative plotting. Fig. S4 (b) shows the spectra before correction, a small shift in peak position can be observed in the enlarged view of the Raman peaks. After correction using a linear pixel shift $S_L$ of -0.6 and a quadratic pixel shift $S_Q$ of -0.4, shown in Fig. S4 (c), the shifts have been compensated.

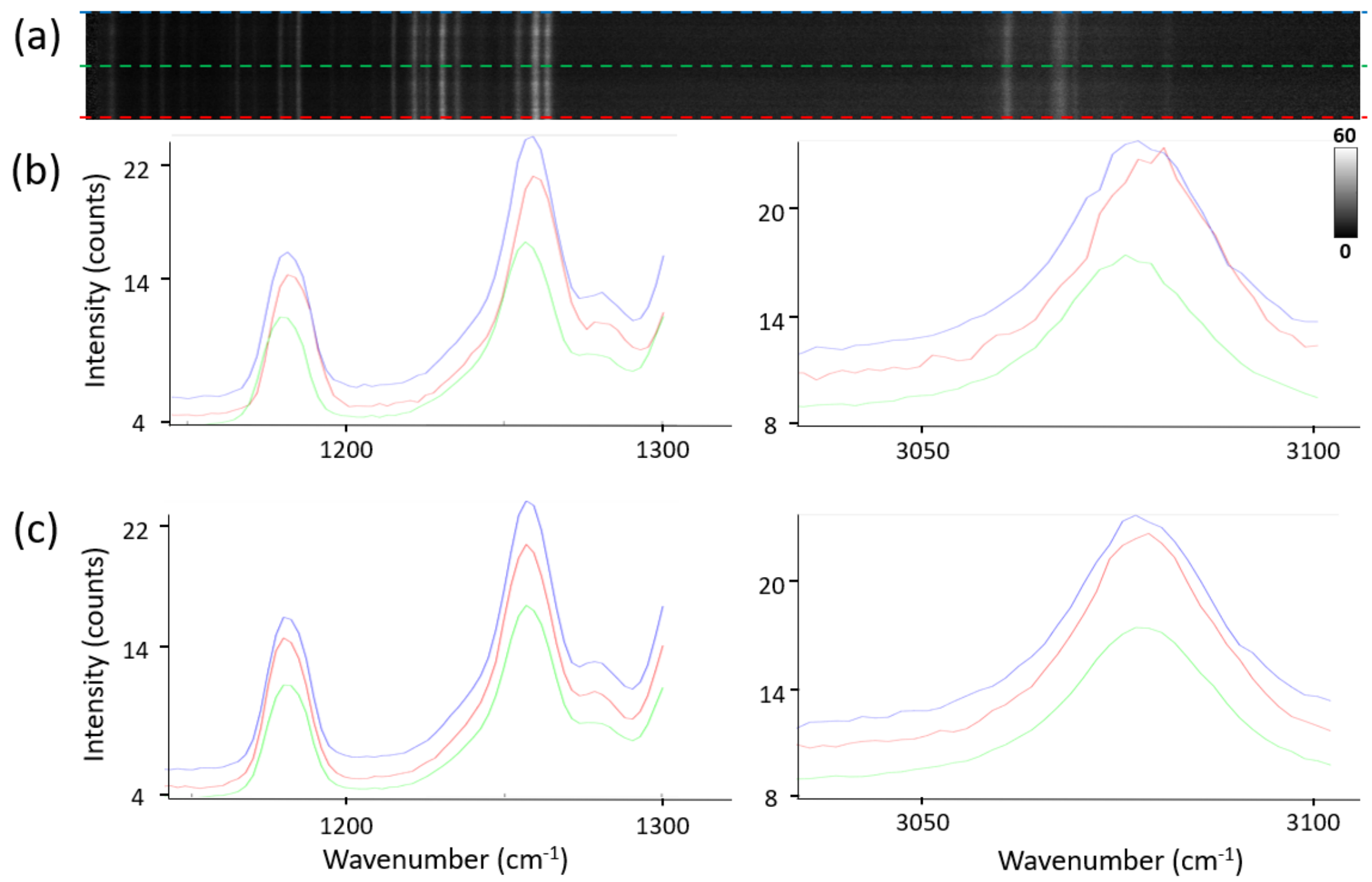

Fig. S4. Spectra before and after correction. (a) greyscale image of a frame acquired on the tablet sample with a resolution of 1920×400, Raman spectra at rows 1 (blue), 200 (green) and 400 (blue) are shown in (b) before correction and in (c) after correction.

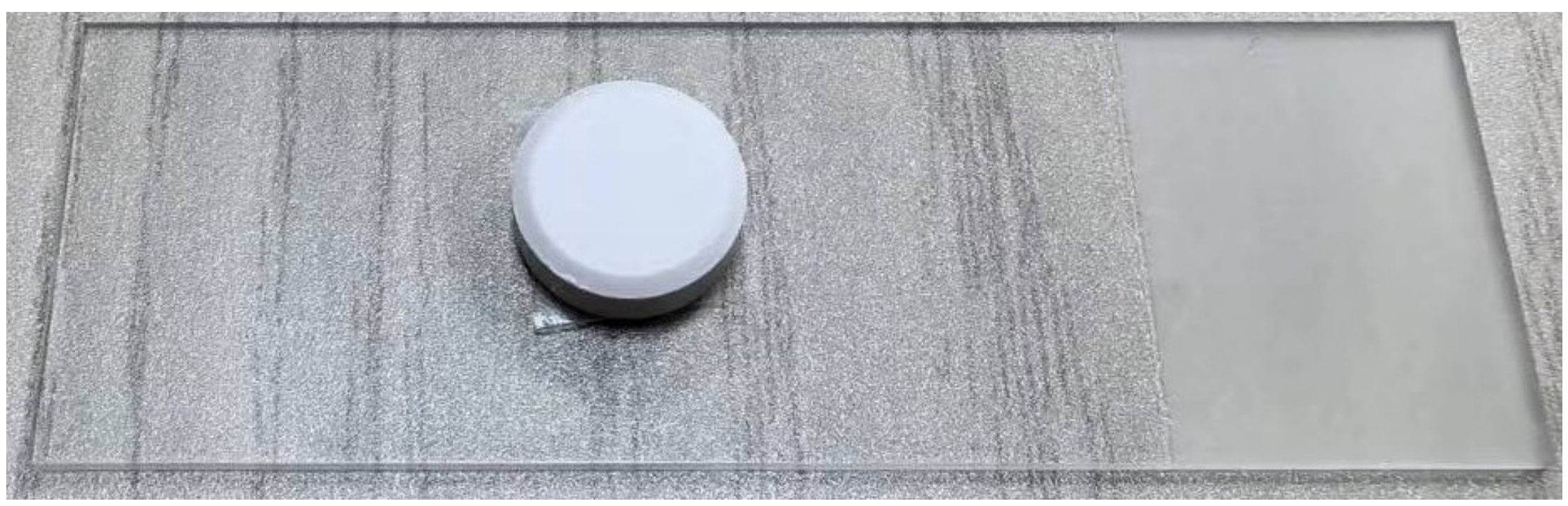

Fig. S5. Photograph of the tablet, secured to the slide using double sided tape

## 4. Spectral calibration

Atomic spectra of neon-argon (Ne-Ar) and mercury (Hg) lamps were collected using the system, and the pixel positions of each characteristic peak were determined using a Gauss-Lorentz fitting method, $f(x; A, \mu, \sigma, \alpha)$, given by

$$f(x; A, \mu, \sigma, \alpha) = \frac{(1-\alpha)A}{\sigma_g\sqrt{2\pi}} e^{\left[-(x-\mu)^2/2\sigma_g{}^2\right]} + \frac{\alpha A}{\pi}\left[\frac{\sigma}{(x-\mu)^2+\sigma^2}\right],$$

where $x$ is the pixel number; $A$ is the peak area; $\mu$ is the peak center position; $\sigma_g$ is the standard deviation of the Gaussian part, and $\sigma$ is the half-width at half maximum of the Lorentzian part; $\alpha$ is a parameter between 0 and 1 defining the relative contribution of the Gaussian and Lorentzian parts. The wavelength range of the signal light received by the spectral detector is 543 nm-666.8 nm, in which the spectral lines of the Ne-Ar and Hg lamps have 18 characteristic peaks and 3 characteristic peaks, respectively, with the wavelengths of the characteristic peaks of the Hg lamps being 546.075 nm, 576.961 nm and 579.067 nm, and the wavelengths of the characteristic peaks of the Ne-Ar lamp being 585.249 nm, 588.190 nm, 594.483 nm, 597.553 nm, 603 nm, 607.434 nm, 609.616 nm, 614.306 nm, 616.359 nm, 621.728 nm, 626.650 nm, 630.479 nm, and 633.443 nm, respectively, 638.299 nm, 640.225 nm, 650.653 nm, 653.288 nm, and 659.895 nm. The Gaussian-Lorentzian fit above was used to establish the peak position information corresponding to each eigen-peak and the standard wavelength information for that peak position. The mathematical relationship between pixel and wavelength was established using the fourth order polynomial

$$\lambda(p) = a_0 + a_1 p + a_2 p^2 + a_3 p^3 + a_4 p^4 ,$$

where $p$ denotes the pixel index, $\lambda(p)$ denotes the wavelength corresponding to that pixel, and $a_0$, $a_1$, $a_2$, $a_3$, and $a_4$ are the coefficients of the polynomial. These coefficients are solved by a fitting process to accurately describe the relation between the pixel position and wavelength, with the resulting fit and parameters shown in Fig. S6. As shown in Fig. S7, the residual is below 5 pm corresponding to 0.07 pixels.

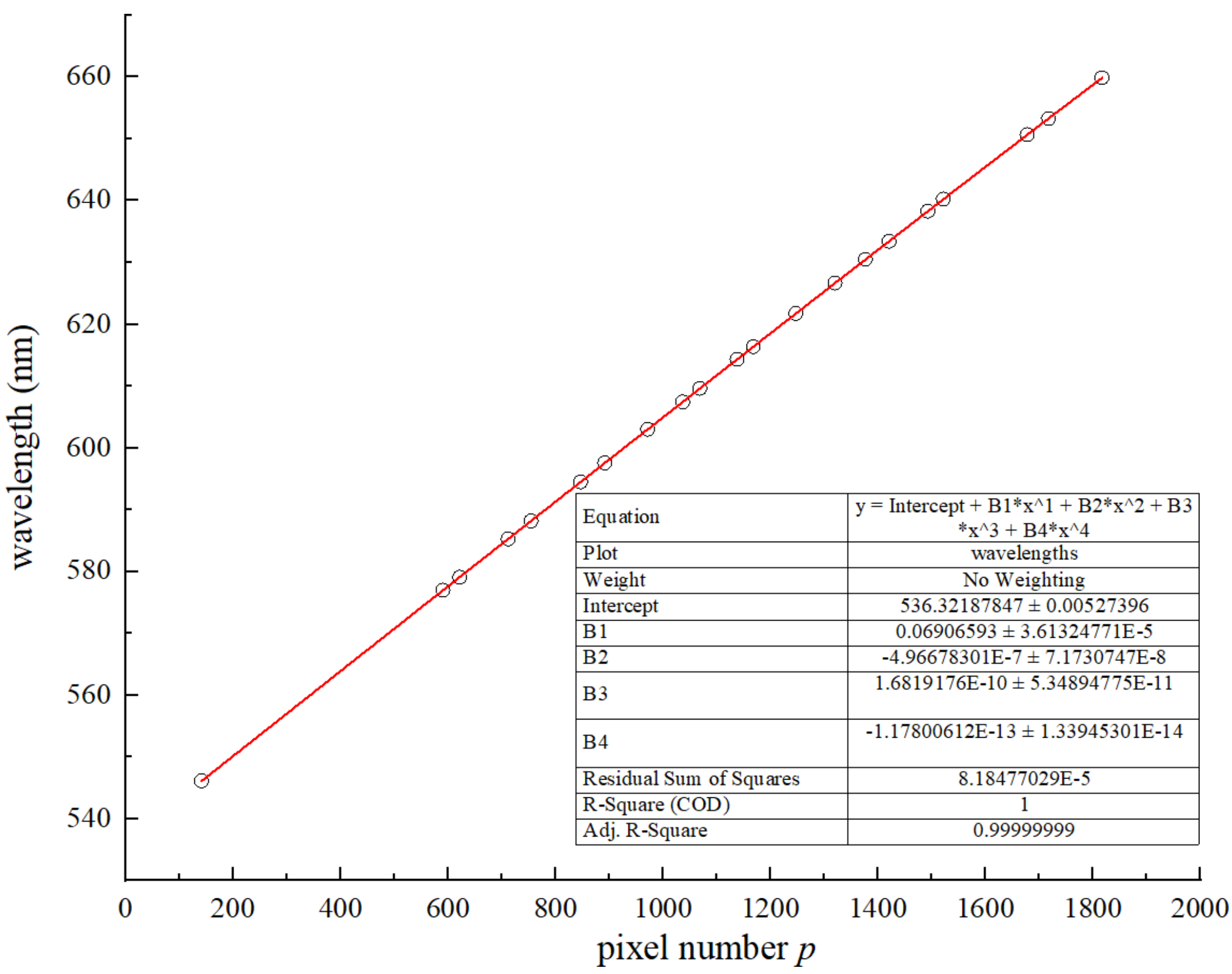

Fig. S6. Spectral calibration – 4th order polynomial fit

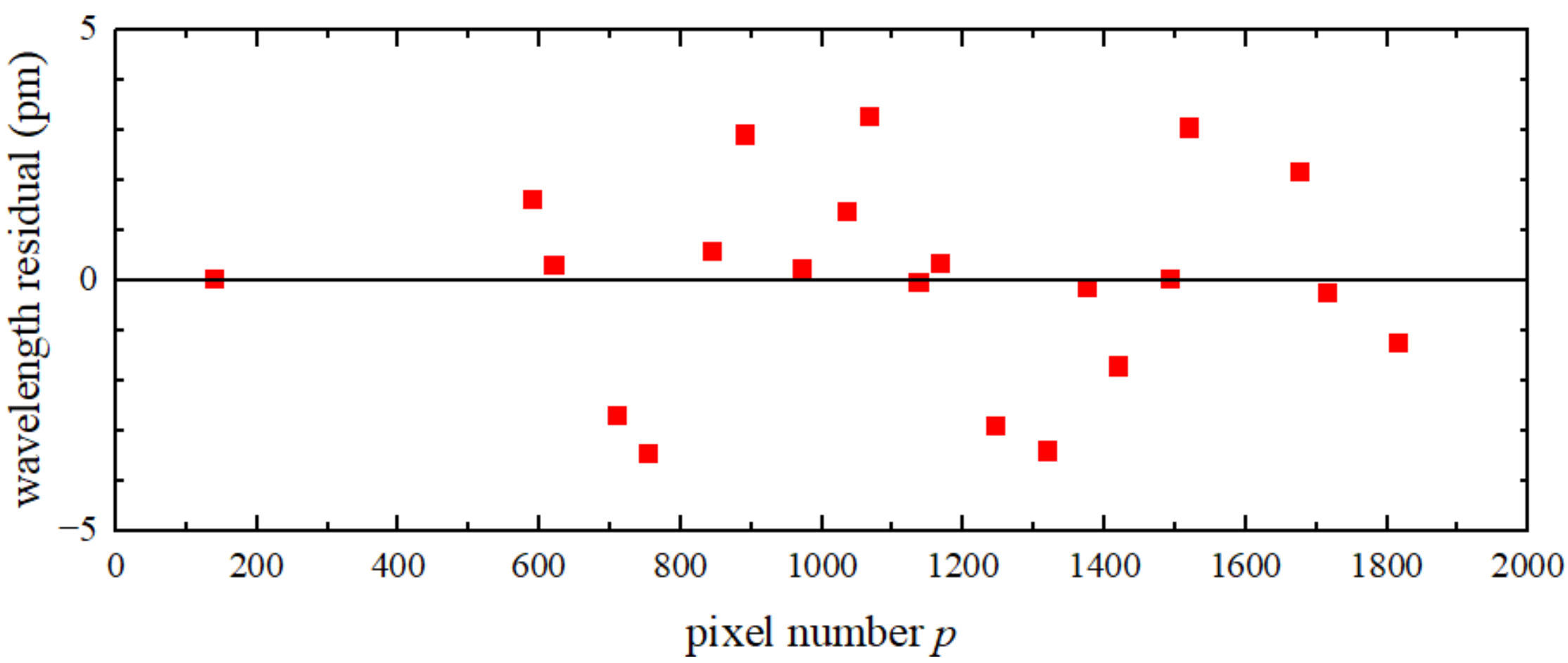

Fig. S7. Residual of the spectral calibration fit

The silicon spectra were collected using the system and fitted to obtain the pixel position of the main silicon Raman peak at 520.7 $cm^{-1}$. Based on the results of the pixel wavelength fitting, the wavelength of the silicon Raman peak at that position, $\lambda_{Si}$, from which the excitation wavelength, $\lambda_L = 532.004$nm was determined using

$$\left(\frac{1}{\lambda_L} - \frac{1}{\lambda_{Si}}\right) = 520.7\,\text{cm}^{-1}.$$

The Raman shift $w(p)$ corresponding to each pixel $p$ is then calculated as

$$w(p) = \left(\frac{1}{\lambda_L} - \frac{1}{\lambda(p)}\right).$$

## 5. Offset and dark noise removal

The offset of a detector is the baseline signal generated by the detector circuitry under no-light conditions, and the dark noise of a detector arising mainly from electronic thermal activity or charge leakage in the absence of illumination, and is proportional to the exposure time. $S_M(\lambda)$ denotes the image data after removal of offset and dark noise. Offset and dark noise are removed by subtracting from the raw data $S(\lambda)$ the dark ambient image data collected at the same integration time $\overline{S_d(\lambda)}$, yielding $S_M(\lambda) = S(\lambda) - \overline{S_d(\lambda)}$.

## 6. Spectral intensity calibration

Using the 532 nm standard reference material NIST, the spectral image data was corrected by obtaining an image intensity correction coefficient matrix from the actual measured NIST spectral image with its standard intensity distribution. Fig. S8 shows the comparison of the results before and after intensity correction on the image data of the NIST standard material, where Fig. S8 (a) shows the original image data after removing the camera offset and dark noise, and Fig. S8 (b) shows the image after intensity correction. Comparing the images it can be clearly observed that the correction process has improved the homogeneity. Fig. S8 (c) shows the comparison of spectra before and after the correction of the 100th row. It can be seen that the intensity of the corrected data in the long-wavelength band portion increases significantly, reflecting the lower quantum efficiency of the camera and the higher dispersion in wavenumbers. The vertical inhomogeneity is also significantly reduced, as shown by the intensity along the 1000th column of the image shown in Fig. S8 (d). The stripes are attributed to inhomogeneities of the spectrometer input slit S2 and of the excitation line.

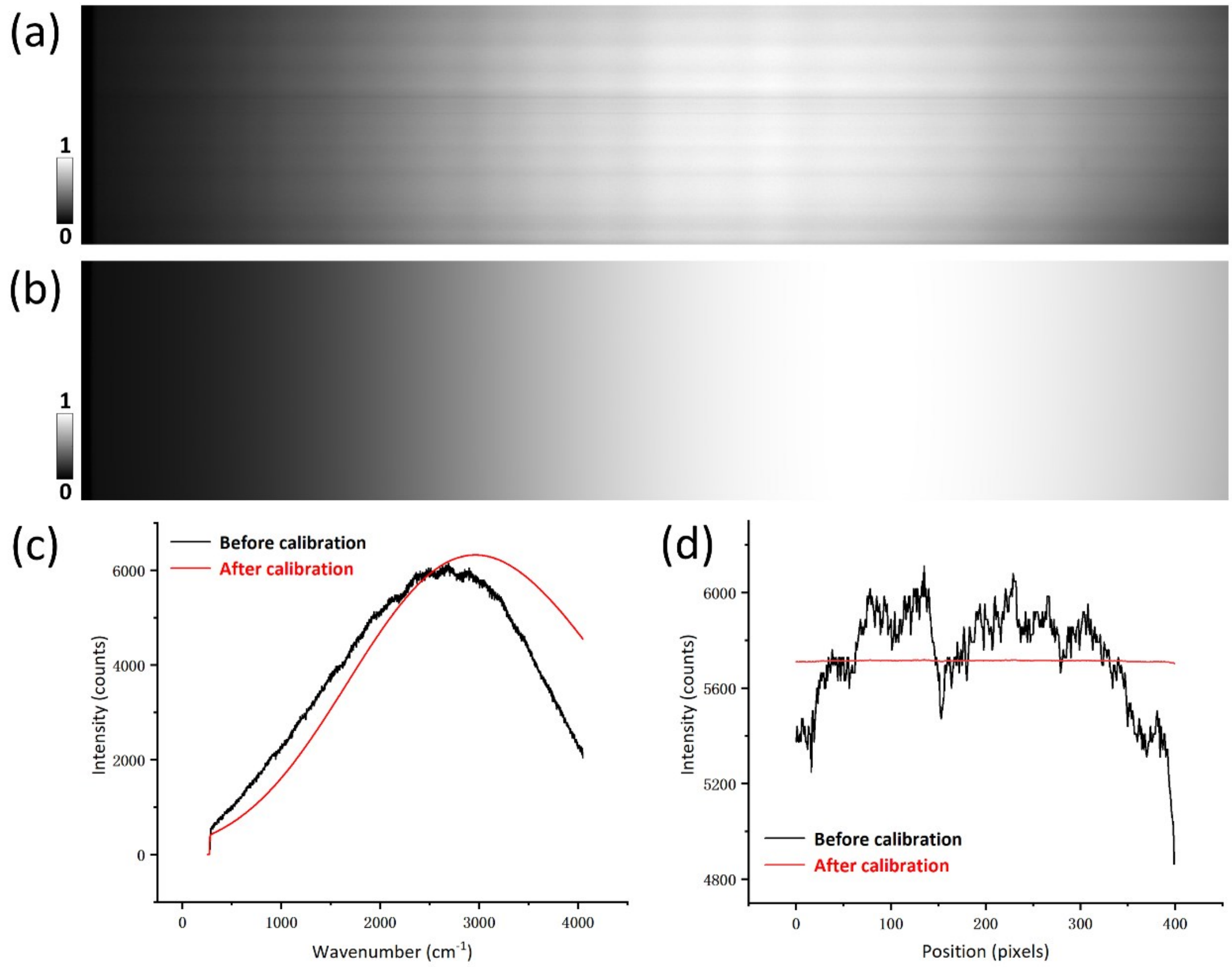

Fig. S8. Comparison of NIST calibration sample images before and after intensity correction. (a) NIST raw image data after removal of offset and dark noise (image size 1920×400 pixels), (b) intensity-corrected results, (c) before-and-after correction spectra for the 100th row of the image, and (d) before-and-after correction intensity for the 1000th column of the image.

**7. Spectral denoising**

Using Singular Value Decomposition (SVD), the spectral data matrix $D$ is decomposed into $D = U\Sigma V^T$ where the matrix $\Sigma$ is a diagonal matrix where the diagonal elements are singular values, the matrix $U$ is the $M \times M$ matrix containing the set of left singular vectors (basis vectors), and the matrix $V$ is the $N \times N$ matrix containing the set of right singular vectors. Larger singular values correspond to more dominant features in the data, while small singular values are usually associated with noise. The reconstructed diagonal matrix $\bar{\Sigma}$ is obtained by retaining only the larger singular values with physically meaningful spectra and setting the other singular values to zero, such that the denoised data $\bar{D}$ is $\bar{D} = U\bar{\Sigma}V^T$.

Through the above process, SVD noise reduction is essentially a process of data compression and reconstruction. Noise reduction is achieved by retaining the main components of the data (larger singular values and their corresponding singular vectors) while removing the components considered as noise (smaller singular values and their corresponding vectors).

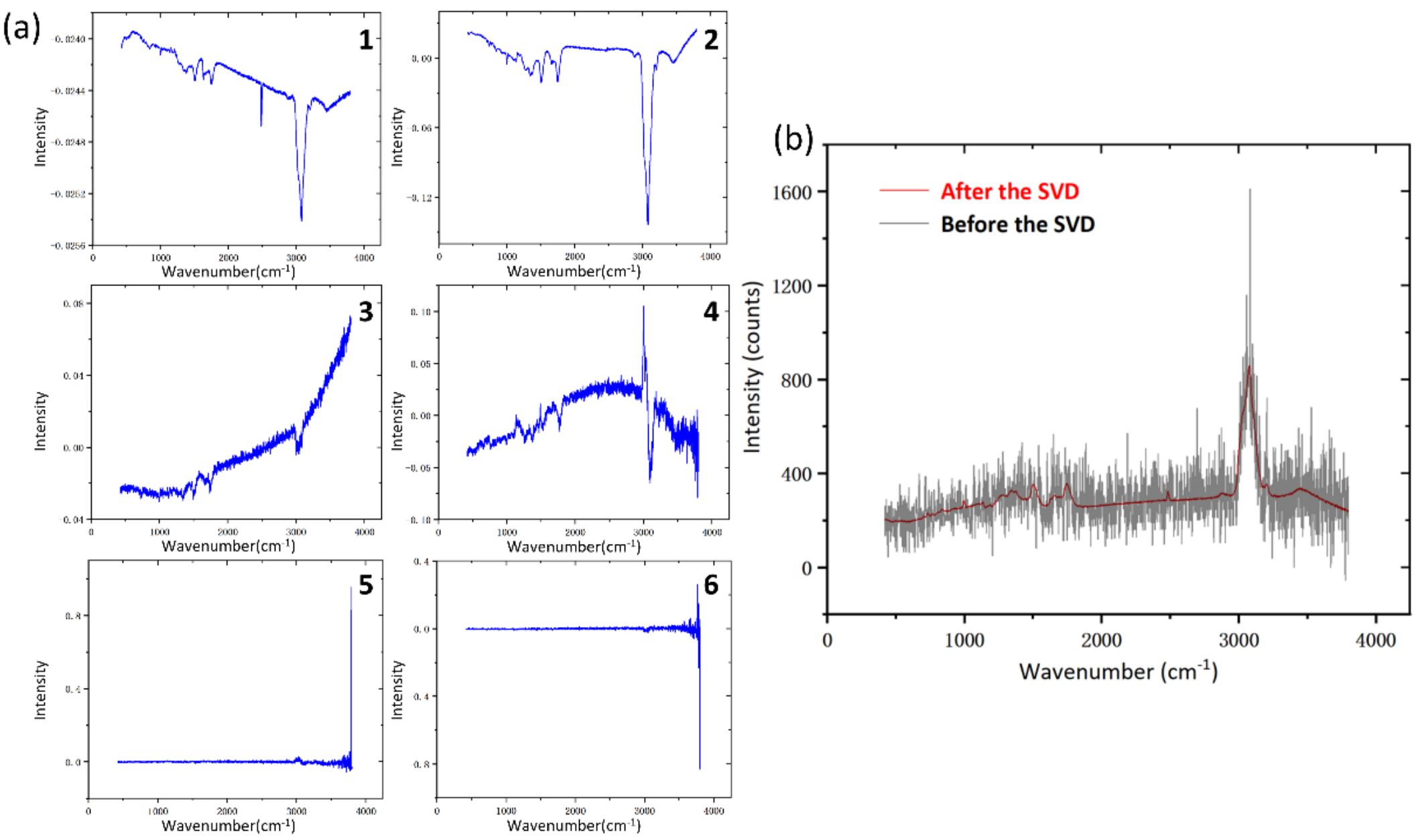

Fig. S9. Raw spectrum and Singular value decomposition. (a) Spectra of the first 6 left singular vectors after SVD decomposition, (b) Comparison of an example spectrum before (grey) and after (red) noise reduction by SVD using the first 4 basis vectors. One count corresponds to 0.084 photoelectrons.

Fig. S9 demonstrates the results of the SVD processing of Raman spectral data of HepG2 cells. It shows a raw spectrum and its SVD decomposition. And this is shown in counts. One count is 1.34e in 8 bit mode, but the data is shown in 12 bit, so a factor of 16 larger in counts. Thus 1 count is 0.084 photoelectrons. (a) displays the spectra of the first 6 left singular vectors. The first 4 vectors were retained during the SVD reconstruction process considering their spectra dominated by Raman-like features. (b) compares the Raman spectral data before and after the SVD noise reduction process, the grey curve provides the unprocessed data, and the red curve indicates the SVD-processed data. It can be clearly seen that through the SVD, noise components in the original data are effectively removed, leaving a smoother and clearer spectral signal.

**8. Comparison of the substrate spectrum with fused silica**

The spectral shape of fused quartz (also called fused silica) from 300 to 600/cm is shown in Fig. S10. The component Fig.2f, reproduced in Fig. S11 bottom left, appears to be a broadened version of this shape. Since the signal is very weak (1 pe maximum), the spectral shape of the component is less defined by the data within the noise, so that the relatively broad Raman spectrum of fused silica, which is considered a fluorescence component, is broadened during the FSC[3] iteration.

The stronger Raman spectra (see Fig. S11) are better constrained by the data and are accordingly not undergoing significant broadening, and the stronger fluorescence components (g,h) are fluorescence from material which appear to have dried around the microplastic.

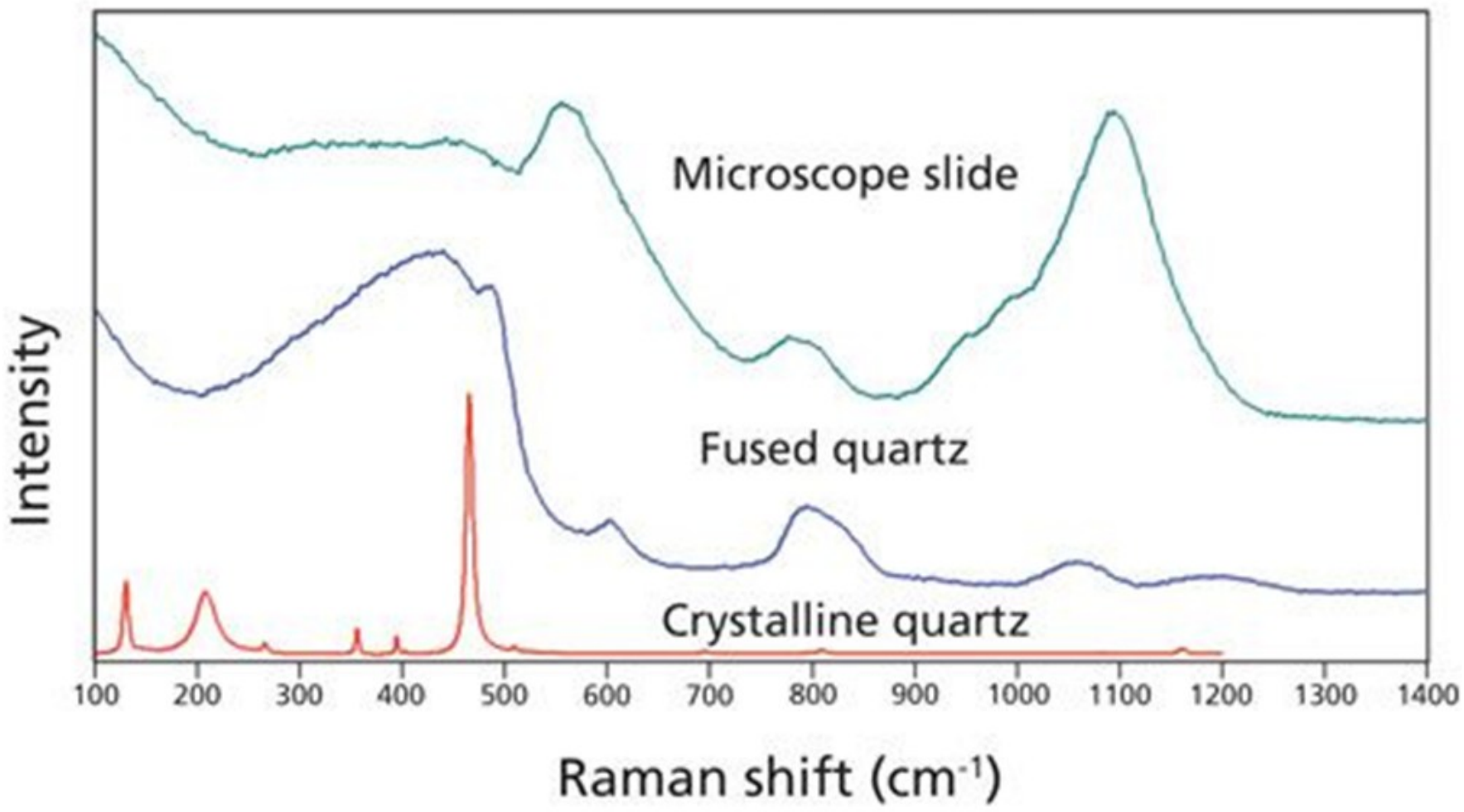

Fig. S10. Raman spectra of crystalline quartz, fused quartz, and a microscope slide reproduced from [Tuschel, David. (2017). Why are the Raman spectra of crystalline and amorphous solids different?. Spectroscopy 32, 3].

## 9. Larger plots of the spectra shown in all figures 2,3,4,5

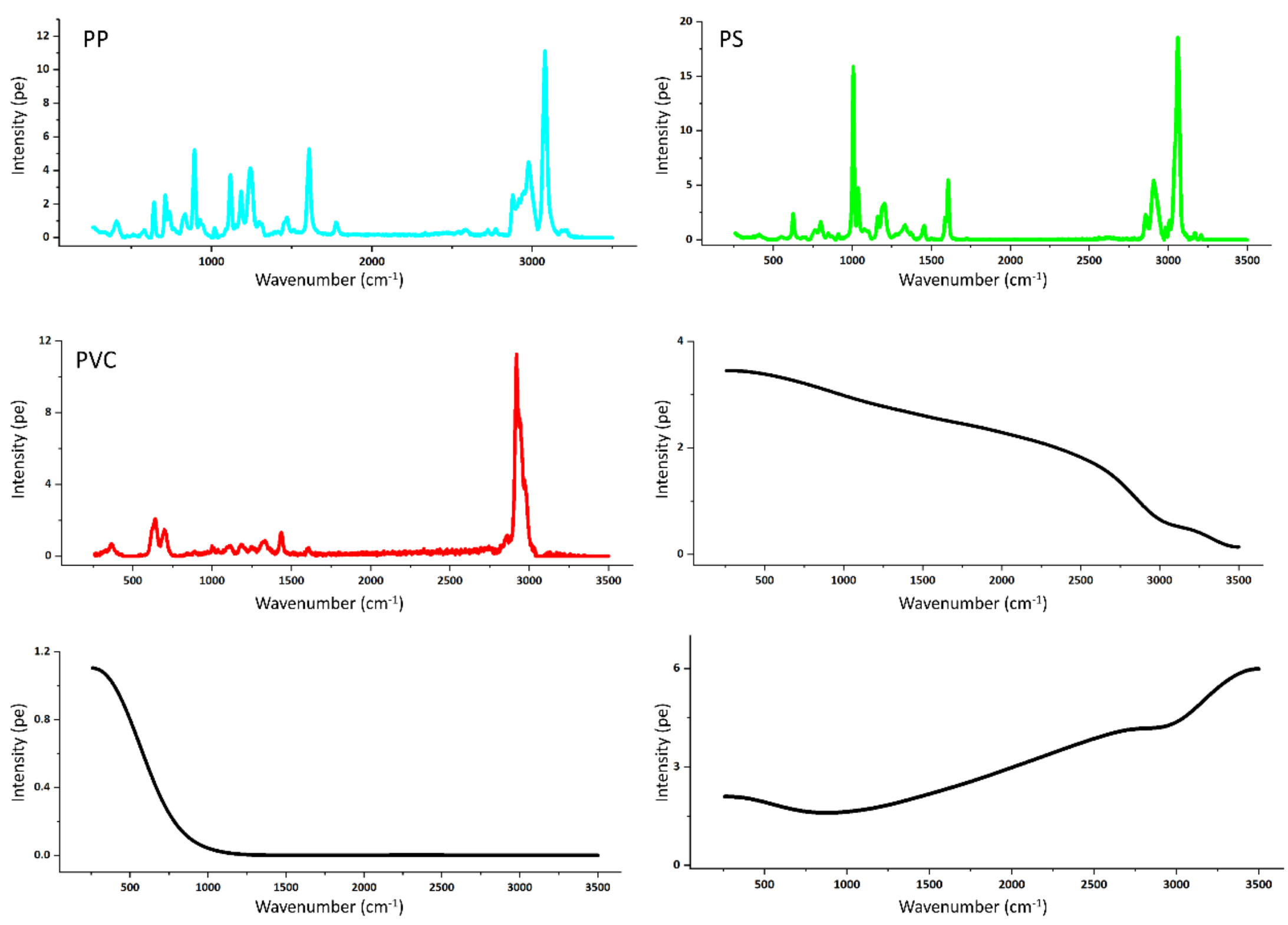

Fig. S11. Larger graphs of the spectra shown in Fig 2 of the main text, provided for clarity.

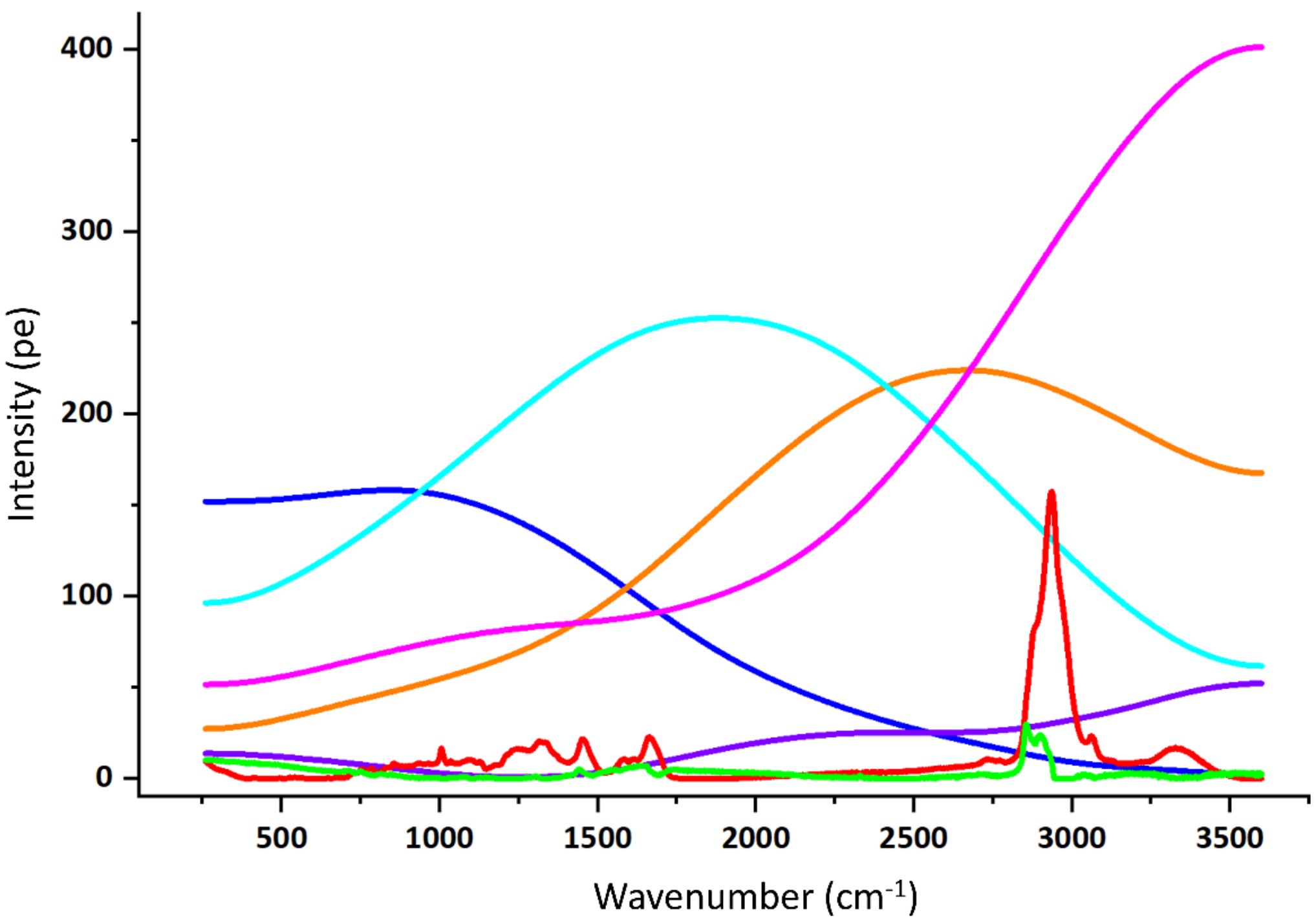

Fig. S12. Larger graphs of the spectra shown in Fig 3 of the main text, provided for clarity.

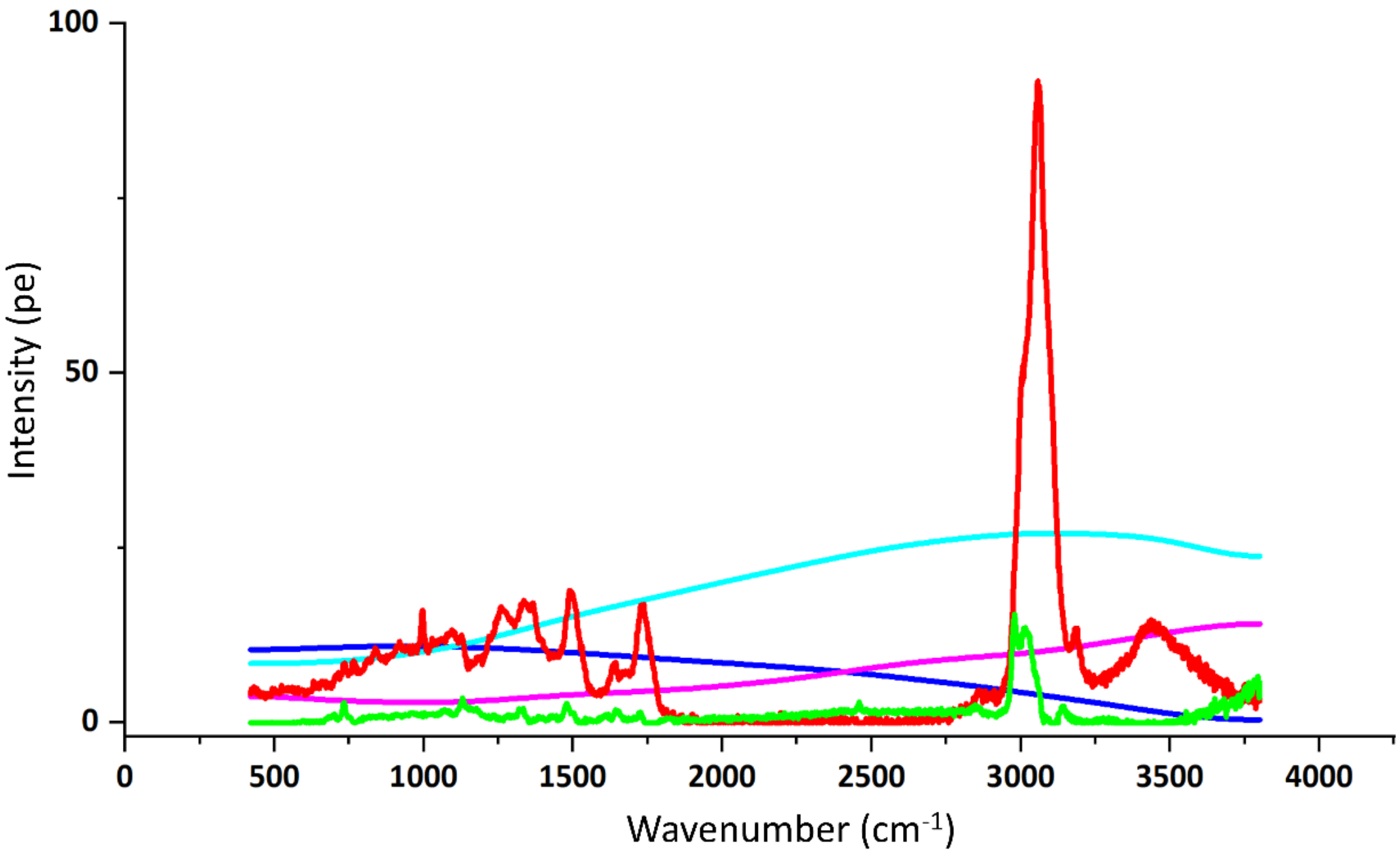

Fig. S13. Larger graphs of the spectra shown in Fig 4 of the main text, provided for clarity.

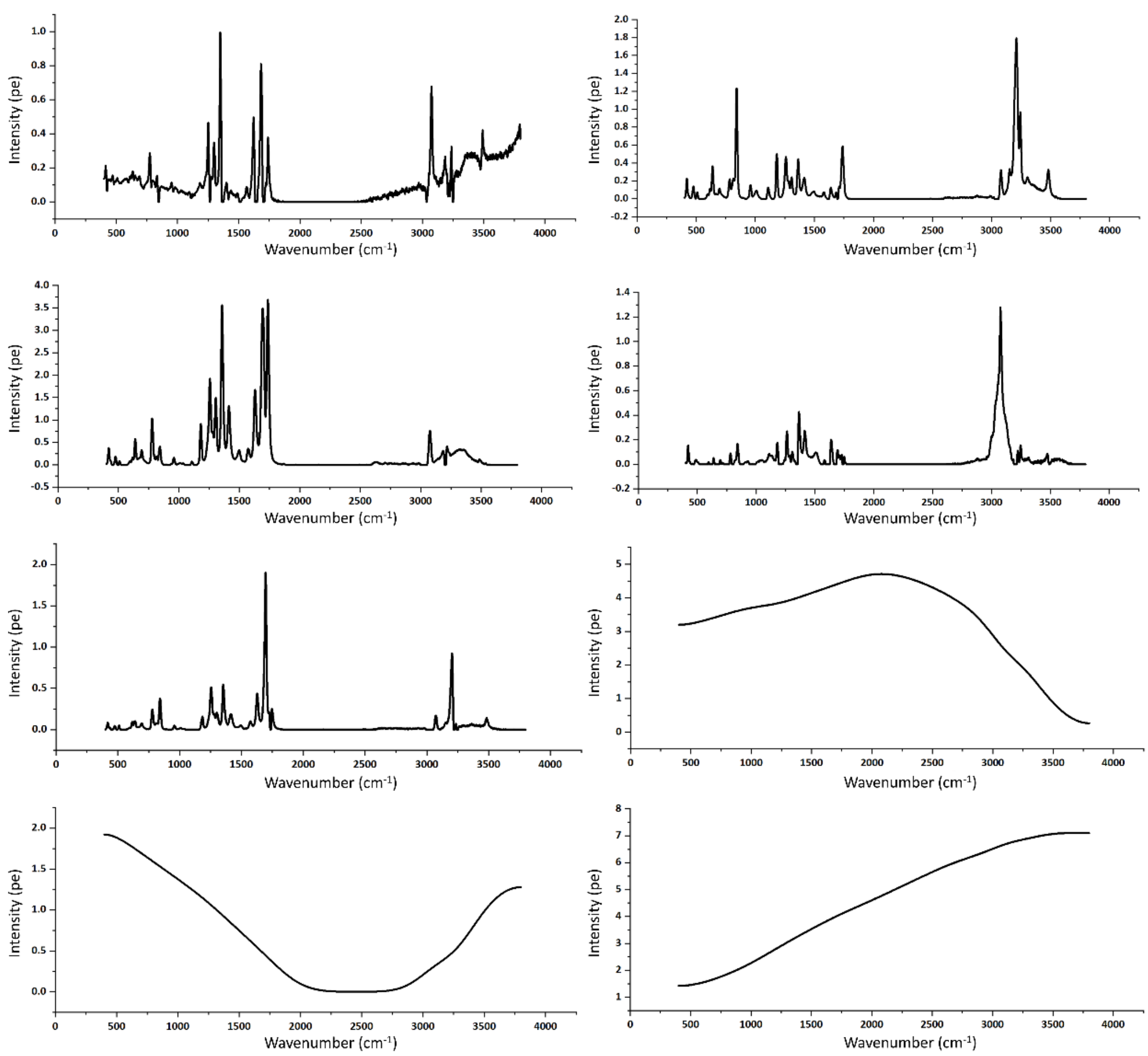

Fig. S14. Larger graphs of the spectra shown in Fig 5 of the main text, provided for clarity.

Table.S1. The full information of tablets given by the supplier.

| Information | Details |
| --- | --- |
| Name of Medicinal Product | Paracetamol Tablets |
| Ingredients | Each tablet contains 0.5 grams of paracetamol. Excipients: starch, thiourea, magnesium stearate, sodium carboxymethyl starch, dextrin. |
| Characteristics | This product is a white tablet. |
| Drug Approval Numbers | Guo Yao Zhun Zi (National Drug Approval Number) H22020911 |

## 10. Data

The original data for the results shown in Fig.2,3,4,5 are provided under the link https://1drv.ms/u/c/b3219e77e76a7d44/IQDM1KSuClV2R6G1_lDytjboAfxMJtuxizrzTJKVvpBf0XA. They contain the raw data as 8bit images as acquired, the SVD denoised data, giving the spectra (text) of the retained components and the spatial concentration maps (16 bit tiff), and the FSC3 analysed data as spatial concentration maps and error maps (16 bit tiff) and the component spectra (text).